\documentclass{svjour3} 

\usepackage{cite}
\usepackage[dvipsnames]{xcolor}
\usepackage{natbib}
\usepackage{verbatim}
\usepackage{caption}
\usepackage{subfigure}
\usepackage{graphicx}
\usepackage{booktabs}
\usepackage{relsize}
\usepackage{colortbl}
\usepackage{enumerate}
\usepackage{amsmath}
\usepackage{tikz}
\usepackage{pgf}
\usepackage{pbox}
\usepackage{rotating}
\usepackage{multirow}
\usepackage{balance}
\usepackage{tablefootnote}
\usepackage{flushend}
\usepackage{natbib}
\usepackage{hyperref}

\usepackage{soul}

\renewcommand{\labelitemi}{--}
\newcommand{\startlist}{\begin{list}{\labelitemi}{\leftmargin=1em}\setlength{\itemsep}{-1mm}}
\newcommand{\stoplist}{\end{list}}

\newcommand{\smallsection}[1]{\noindent {\bf \underline{#1}}.\hspace{1mm}}
\newcommand{\ea}{{\em et al.}}

\usepackage{algorithmicx}
\usepackage{algorithm} 
\usepackage{algpseudocode}
\usepackage{graphicx}
\usepackage{wrapfig}

\usepackage{tikz}
\newcommand*\circled[1]{\tikz[baseline=(char.base)]{
            \node[shape=circle,draw,inner sep=0.2pt] (char) {#1};}}

\newcommand{\ourtool}{\textsc{AIBugHunter}}

\definecolor{mygreen}{rgb}{0.0, 0.5, 0.0}

\newcommand{\revisedinline}[1]{{\color{black}{#1}}}

\newcommand{\rqone}{How accurate is our approach for predicting vulnerability IDs (i.e., CWE-IDs)?}
\newcommand{\rqtwo}{How accurate is our approach for predicting vulnerability types (i.e., CWE abstract types)?}
\newcommand{\rqthree}{How accurate is our approach for predicting vulnerability severity?}
\newcommand{\rqfour}{How do the software practitioners perceive the usefulness of our \ourtool?}

\journalname{}

\begin{document}

\title{\ourtool: A Practical Tool for Predicting, Classifying and Repairing Software Vulnerabilities}



\author{Michael Fu \and \\
        Chakkrit Tantithamthavorn \and \\
        Trung Le \and \\
        Yuki Kume \and \\
        Van Nguyen \and \\
        Dinh Phung \and \\
        John Grundy
}


\institute{ Michael Fu \at
              Monash University, Clayton, VIC, Australia \\
              \email{yeh.fu@monash.edu}  \\
            \\
            Chakkrit Tantithamthavorn \at
              Monash University, Clayton, VIC, Australia \\
              \email{chakkrit@monash.edu}  \\
            \\ 
            Trung Le \at
              Monash University, Clayton, VIC, Australia \\
              \email{trunglm@monash.edu} \\
            \\
            Yuki Kume \at
              Monash University, Clayton, VIC, Australia \\
              \email{yuki.kume@monash.edu} \\
            \\
            Van Nguyen \at
              Monash University, Clayton, VIC, Australia \\
              \email{Van.Nguyen1@monash.edu} \\
            \\
            Dinh Phung \at
              Monash University, Clayton, VIC, Australia \\
              \email{Dinh.Phung@monash.edu} \\
            \\
            John Grundy \at
              Monash University, Clayton, VIC, Australia \\
              \email{John.Grundy@monash.edu} \\              
}

\date{Received: date / Accepted: date}

\maketitle

\begin{abstract} 
Many Machine Learning(ML)-based approaches have been proposed to automatically detect, localize, and repair software vulnerabilities.
While ML-based methods are more effective than program analysis-based vulnerability analysis tools, few have been integrated into modern Integrated Development Environments (IDEs), hindering practical adoption.
\revisedinline{To bridge this critical gap, we propose in this article \ourtool, a novel Machine Learning-based software vulnerability analysis tool for C/C++ languages that is integrated into the Visual Studio Code (VS Code) IDE.
\ourtool~helps software developers to achieve real-time vulnerability detection, explanation, and repairs during programming.
In particular, \ourtool~scans through developers' source code to (1) locate vulnerabilities, (2) identify vulnerability types, (3) estimate vulnerability severity, and (4) suggest vulnerability repairs.
We integrate our previous works (i.e., LineVul and VulRepair) to achieve vulnerability localization and repairs.
In this article, we propose a novel multi-objective optimization (MOO)-based vulnerability classification approach and a transformer-based estimation approach to help \ourtool~accurately identify vulnerability types and estimate severity.
Our empirical experiments on a large dataset consisting of 188K+ C/C++ functions confirm that our proposed approaches are more accurate than other state-of-the-art baseline methods for vulnerability classification and estimation.
Furthermore, we conduct qualitative evaluations including a survey study and a user study to obtain software practitioners’ perceptions of our \ourtool~tool and assess the impact that \ourtool~may have on developers’ productivity in security aspects.
Our survey study shows that our \ourtool~is perceived as useful where 90\% of the participants consider adopting our \ourtool~during their software development.
Last but not least, our user study shows that our \ourtool~could possibly enhance developers’ productivity in combating cybersecurity issues during software development.}
\ourtool~is now publicly available in the Visual Studio Code marketplace.

\end{abstract}

\section{Introduction}
Software vulnerabilities are weaknesses in an information system, security procedures, internal controls, or implementations that could be exploited or triggered by a threat source~\cite{johnson2011guide}.
Such unresolved weaknesses result in extreme security or privacy risks.
According to the research conducted by~\cite{mend} on open source vulnerabilities in the past 10 years (including multiple sources like the National Vulnerability Database (NVD), security advisories, GitHub issue trackers etc.), C has the highest number of vulnerabilities out of all seven reported languages (i.e., C, PHP, Java, JavaScript, Python, C++, Ruby), accounting for 47\% of all reported vulnerabilities.
Buffer errors (e.g., CWE-119: Improper Restriction of Operations within the Bounds of a Memory Buffer) are the most common vulnerability in C and C++.
It is worth noting that this group of vulnerabilities related to memory corruption could often have critical consequences such as system crashes and sensitive information disclosure.
In particular, our proposed software vulnerability classification approach can correctly identify 79\% of the CWE-119 buffer error as shown in Table~\ref{tab:movul_cwe_list} (see Rank 17).


Recently, the shift-left testing concept (i.e. move software testing earlier in project timelines) has been proposed to try to perform software testing at earlier stages of development, instead of testing applications during late phases of development.
Thus, vulnerabilities related to fundamental features, such as buffer errors, could ideally be found and fixed earlier.
DevSecOps has also been proposed to extend the idea of DevOps by integrating security into DevOps initiatives~\citep{devsecops}.
DevSecOps aims to examine application security from the start of development by automating some security gates and selecting the right tools to continuously integrate security in the DevOps workflow. 
For example,
program analysis(PA)-based tools can be integrated into IDEs, such as Visual Studio Code (VS Code), to detect such vulnerabilities during coding.
However, these methods usually rely on pre-defined vulnerability patterns and struggle to detect specific types of vulnerability.
\cite{croft2021empirical} demonstrated that Machine Learning(ML)-based techniques are more accurate than PA-based tools in detecting file-level vulnerabilities.
Our own previous study showed that our ML-based LineVul approach is more accurate than the PA-based Cppcheck tool~\citep{cppcheck} on line-level vulnerability prediction~\citep{fu2022linevul, pornprasit2021jitline, pornprasit2022deeplinedp}.
ML-based methods learn vulnerability patterns based on historical vulnerability data instead of relying on pre-defined patterns. Thus, ML-based approaches can capture more kinds of vulnerabilities and be more easily extended as new vulnerabilities emerge. 
PA-based tools, such as Checkmarx~\citep{checkmarx}, have been integrated into software development workflow to support security diagnosis during development. However, to date, ML-based tools have not been integrated as security tools to help detect security issues during software development. 

In this article, we propose an ML-based software vulnerability analysis tool, \ourtool, to bridge the critical gap between ML-based security tools and software practitioners.
\ourtool~is integrated into a modern IDE (i.e., VS Code) -- to fulfil the concept of shift-left testing and to support real-time security inspection during software development.
In particular, given developers' source code written in C/C++, our \ourtool~can (1) locate vulnerabilities, (2) classify vulnerability types, (3) estimate vulnerability severity, and (4) suggest repairs.
We integrate our previous work LineVul~\citep{fu2022linevul} and VulRepair~\citep{fu2022vulrepair} for \ourtool~to achieve automated vulnerability localization and repairs.
In this article, we further propose a multi-objective optimization (MOO)-based approach to optimize the multi-task learning scenario and help our \ourtool~accurately identify vulnerability types (i.e., CWE-IDs, and CWE-Types) and explain the detected vulnerabilities.
In addition, a transformer-based approach is proposed to help \ourtool~estimate the vulnerability severity (i.e., CVSS Score) which could be beneficial for the prioritization of security issues.


We evaluate our proposed MOO-based vulnerability classification and severity estimation approaches on a large dataset that consists of 188k+ C/C++ functions including various vulnerability types and severity.
We found that our MOO-based vulnerability classification approach outperforms other baseline methods and achieves the accuracy of 65\% (demonstrated in RQ1) and 74\% (demonstrated in RQ2) for classifying CWE-ID and CWE-Types respectively. In addition, our transformer-based severity estimation approach outperforms other baseline methods and achieves the best mean squared error (MSE) and mean absolute error (MAE) measures (demonstrated in RQ3).
We evaluate our \ourtool~through qualitative evaluations including (1) a survey study to obtain software practitioners' perceptions of our \ourtool~tool; and (2) a user study to investigate the impact that our \ourtool~could have on developers’ productivity in security aspects.
Our survey study shows that predictions provided by AIBugHunter are perceived as useful by 47\%-86\% of participated software practitioners and 90\% of participants will consider adopting our \ourtool.
Moreover, our user study indicates that \ourtool~could save developers' time spent on security analysis that could potentially enhance security productivity during software development (demonstrated in RQ4).

The main contributions of this work include:
\begin{enumerate}
    \item \ourtool, a novel ML-based software security tool for C/C++ that is integrated into the VS Code IDE to bridge the gap between ML-based vulnerability prediction techniques and software developers and achieve real-time security inspection;
    \item A quantitative evaluation of \ourtool~on a large dataset showing its high precision and recall;
    \item A qualitative survey study of \ourtool~with 21 software practitioners demonstrating both its practicality and potential acceptance;
    \item A qualitative user study of \ourtool~with 6 software practitioners demonstrating \ourtool~could enhance practitioners’ productivity in combating security issues during software development;
    \item A multi-objective optimization approach for vulnerability classification that optimizes the multi-task learning scenario for classifying the vulnerability types for vulnerable functions written in C/C++; and
    \item A transformer-based approach to estimate vulnerability severity for vulnerable functions written in C/C++.
\end{enumerate}

We make available our datasets, scripts including data processing, model training, model evaluation, and experimental results related to our approach in a GitHub repository: (\url{https://github.com/awsm-research/AIBugHunter}).
Additionally, \ourtool~is available at VS Code marketplace (\url{https://marketplace.visualstudio.com/items?itemName=AIBugHunter.aibughunter}).

The rest of this article is organized as follows. Section~\ref{sec:aibughunter} presents a high-level overview of our \ourtool.
Section \ref{sec:our_approach} presents our approach to predicting vulnerability types and severity.
Section~\ref{sec:experiment} presents our studied datasets, our experimental setup, and our first three research questions along with their results.
Section~\ref{sec:survey} presents a qualitative evaluation of \ourtool~including a survey study and a user study to answer the last research question.
Section~\ref{sec:threats} discloses the threats to validity.
Section~\ref{sec:related} discusses the related works.
Section~\ref{sec:conclusion} draws the conclusions.
\section{\ourtool: Our Approach}
\label{sec:aibughunter}

We provide an overview of our \ourtool, an ML-based vulnerability prediction tool as a plug-in in Visual Studio Code (VS Code).
The main purpose of our \ourtool~is to bridge the gap between ML-based vulnerability prediction techniques and software developers by providing a security plug-in in IDE to present more security information during software development.

\begin{figure}[t]
\includegraphics[width=\linewidth,height=4.5cm]{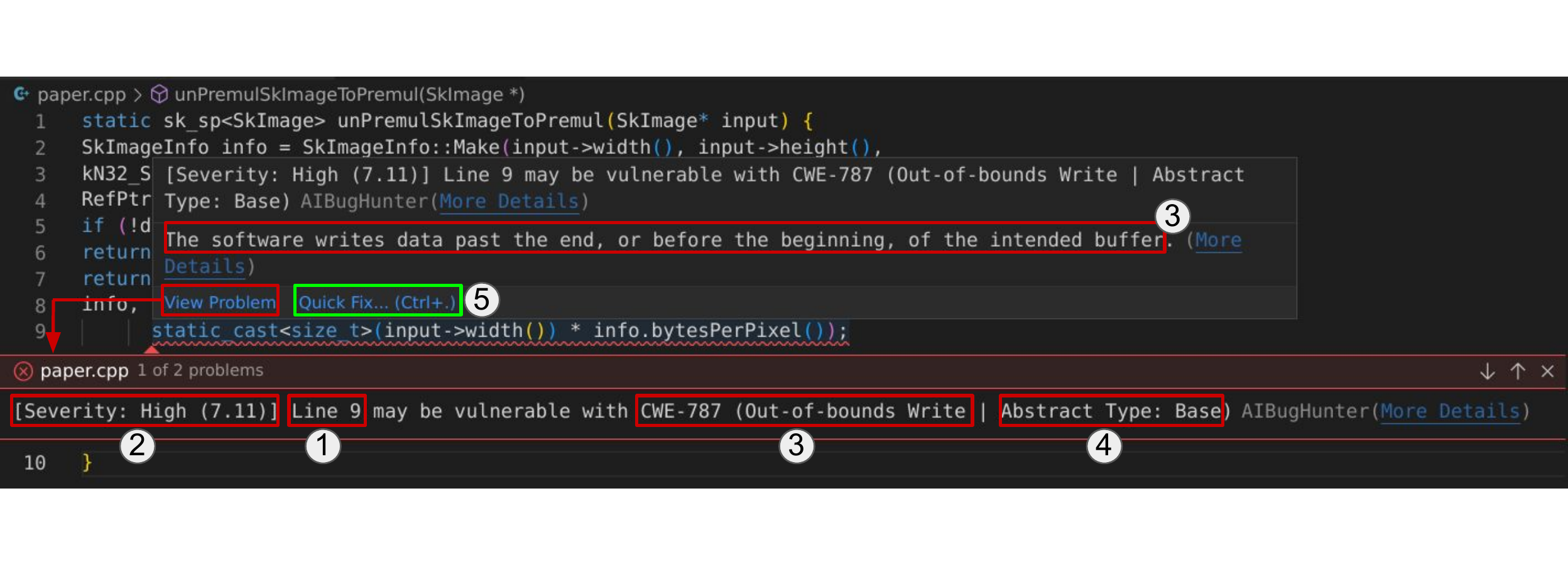}
\caption{The user interface of our \ourtool.}
\label{fig:aib_ui}
\end{figure}

\subsection{\ourtool~security tool}
As a security tool integrated into VS Code,
\ourtool~first scans the file opened by developers and parse the whole file into multiple separate functions.
For each function, our \ourtool~performs the following 4 steps:
\begin{enumerate}
    \item \underline{\emph{Localize}} the vulnerable lines (LineVul);
    \item \underline{\emph{Classify}} the vulnerability types (proposed in this paper); 
    \item \underline{\emph{Estimate}} the vulnerability severity (proposed in this paper); and
    \item \underline{\emph{Suggest}} the repair patches (VulRepair).
\end{enumerate}
where LineVul~\citep{fu2022linevul} locates vulnerable lines;
our approach predicts types and severity;
and VulRepair~\citep{fu2022vulrepair} suggests repairs.

In AIBugHunter, we use LineVul and VulRepair from our previous works. These models were trained using the extensive Big-Vul dataset offered by~\cite{fan2020ac} and the CVEFixes dataset provided by~\cite{bhandari2021cvefixes}. We illustrate both of them as follows:

LineVul is among the first to predict line-level vulnerabilities using the transformer model and its self-attention mechanism.
Given a C/C++ function as input, first, LineVul leverages a BPE tokenizer to tokenize the function into subword tokens and mitigate the out-of-vocab problem.
Second, LineVul leverages transformer encoders~\citep{vaswani2017attention} to learn the representation of those tokens, which can better tackle the long-term dependencies among tokens than previously proposed RNN-based methods~\citep{li2021vulnerability}.
Third, LineVul uses a linear classification head to predict function-level vulnerability prediction based on the learned representations.
LineVul uses intrinsic model interpretation to localize line-level vulnerabilities.
In particular, LineVul summarizes the self-attention scores of each line in the function and ranks the line scores to place potentially vulnerable lines on the top.
Our previous work~\citep{fu2022linevul} has demonstrated that LineVul achieves the best accuracy for both function-level and line-level vulnerability prediction and is the most cost-effective approach to localize line-level vulnerabilities when compared with other baseline methods.

VulRepair is among the first to leverage a large pre-trained language model for the automated vulnerability repair (AVR) problem.
Given a vulnerable C/C++ function as input, instead of using word-level tokenization as previous work~\citep{chen2021neural}, VulRepair leverages a BPE tokenizer to tokenize the function into subword tokens and address the potential OOV problem.
VulRepair uses a pre-trained encoder-decoder T5 architecture where encoders encode the representation of the vulnerable function and decoders generate the corresponding repair patches.
In particular, the relative position encoding of T5 used by VulRepair improves the absolute position encoding of the vanilla transformer used in previous work~\citep{chen2021neural}.
VulRepair was evaluated using the human-written repairs as ground-truth labels where a repair generated by VulRepair is considered correct if it is identical to the labels.
Our previous work~\citep{fu2022vulrepair} has demonstrated that VulRepair substantially improves the performance of previous works for the AVR problem.

\subsection{Example Usage}

Consider the situation where an opened file contains one function written in C++, shown in Fig~\ref{fig:aib_ui}.
This example uses a real-world "out-of-bounds write" vulnerability~\citep{cwe787} that is considered the most dangerous vulnerability in 2021~\citep{cwetop25}.
Fig~\ref{fig:aib_ui} shows
the "\textit{unPremulSkImageToPremul}" function. \ourtool has analyzed this and considered it as a vulnerable function. This is due to the variable type "size\_t" being misused, causing an "out-of-bounds write" vulnerability (i.e., CWE-787) at line number 9.

As shown in Fig~\ref{fig:aib_ui},
\ourtool~first takes the whole function as an input and sends it to its backend models,
LineVul~\citep{fu2022linevul}. The LineVul algorithm identifies that the 9th line of the "\textit{unPremulSkImageToPremul}" function is a vulnerable line, as annotated by \circled{1}.
Our approach further classifies this function as a vulnerability of CWE-787, shown as \circled{2}, with a Base type shown in \circled{3}.

This function is predicted as being of a high severity with a CVSS score of 7. This is shown as \circled{4}.
Finally, we use our backend tool, VulRepair~\citep{fu2022vulrepair}, to generate repair patches. This patch will be used to replace the vulnerable line. The developer can select this option by clicking on the "Quick Fix" button, shown as \circled{5}.

\subsection{\ourtool~Implementation}
We developed our \ourtool~extension using the VS Code Extension API provided by Microsoft to gain symbol information and utilize other VS Code IDE features.
\ourtool~is mainly written in TypeScript following the boilerplates provided by the VS Code extension generator.
Being a plain VS Code extension, our package's operations are driven by a Node.js engine.
In what follows, we introduce the front-end and back-end implementation details of \ourtool.

\subsubsection{Front-End Implementation}
The UI elements of \ourtool~are defined by the VSCode API provided by Microsoft, the backend of which is Node.JS, and is interacted using the TypeScript language.
When a user opens a C/C++ file, \ourtool~extracts each function from the source code using the ``symbols'' information available through VSCode API, and builds a list of parsed functions to be passed into the DL models introduced in the following section.
The back-end will return the generated predictions using the API provided, and the relevant information is displayed on the UI as ``diagnostics''.
This enables the extension to indicate the specific line to fix using underlines, display hover messages, provide a link to the CWE page, provide a ``Quick Fix'' button for repair candidates, and offer other error messages in the interface.
\ourtool~presents its vulnerability predictions and explanatory information as shown in Fig~\ref{fig:aib_ui}.

\begin{figure}[t]
\includegraphics[width=\linewidth,height=4.5cm]{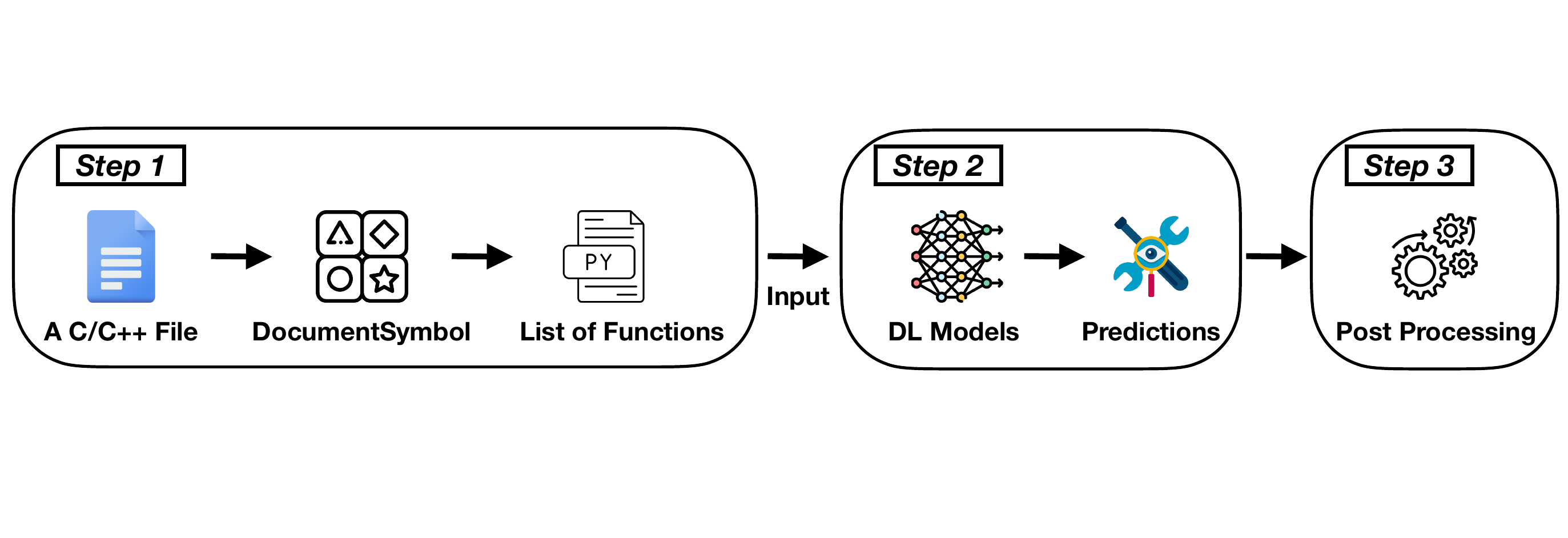}
\caption{The back-end implementation of our \ourtool.}
\label{fig:aib_backend}
\end{figure}

\subsubsection{Back-End Implementation}
The back end consists of three main steps as summarized in Fig~\ref{fig:aib_backend}.
First, the data preparation step to construct data for DL models.
Second, the DL models inference step for (1) locating line-level vulnerabilities, (2) classifying vulnerability types, (3) estimating vulnerability severity, and (4) suggesting repairs.
Third, the post-prediction processing step is used to prepare information and present it in the UI.

\noindent\textbf{Step 1: Data Preparation.}
When a C/C++ file is opened, VSCode automatically analyzes it and generates a ``DocumentSymbol'', which is a collection of symbols in the document such as variables, classes, and functions. We preserve only the collection of functions to construct a list of functions parsed from the document, where each parsed function undergoes formatting to remove comments. Note that all the modifications are recorded as a position delta to correctly map the prediction results to the original code.

\noindent\textbf{Step 2: DL Model Inference.}
The model inference consists of two steps to obtain all the predictions to present in the front end as described below:

\textbf{Step 2a.} Send the list of functions from the data preparation step to the line-level vulnerability detection model's inference API endpoint (or flag in local inference mode). This will return a JSON which tells if individual functions are vulnerable or not (binary), and scores on each line of the function that determines which line the modifications are required to fix the vulnerability.

\textbf{Step 2b.} For functions that were predicted vulnerable in the previous step are now sent to three additional DL models. For each function, the first model will return a CWE-ID indicating the vulnerability type; the second model returns a CVSS score indicating the severity; and the third model returns an annotated piece of ``patch code'' as suggested repairs.

\noindent\textbf{Step 3: Post-Prediction Processing.}
All the predictions from the model inference step are processed according to the user configuration.
Additionally for functions predicted as vulnerable, we fetch the vulnerability description from MITRE ATT\&CK~\cite{mitre} based on the predictions to provide in-depth details of the predicted vulnerability and an accessible link to the official page of the specific CWE-ID. 
Finally, the organized information is displayed on the interface via the VSCode extension API.

\begin{figure*}[t]
\includegraphics[width=\linewidth,height=5cm]{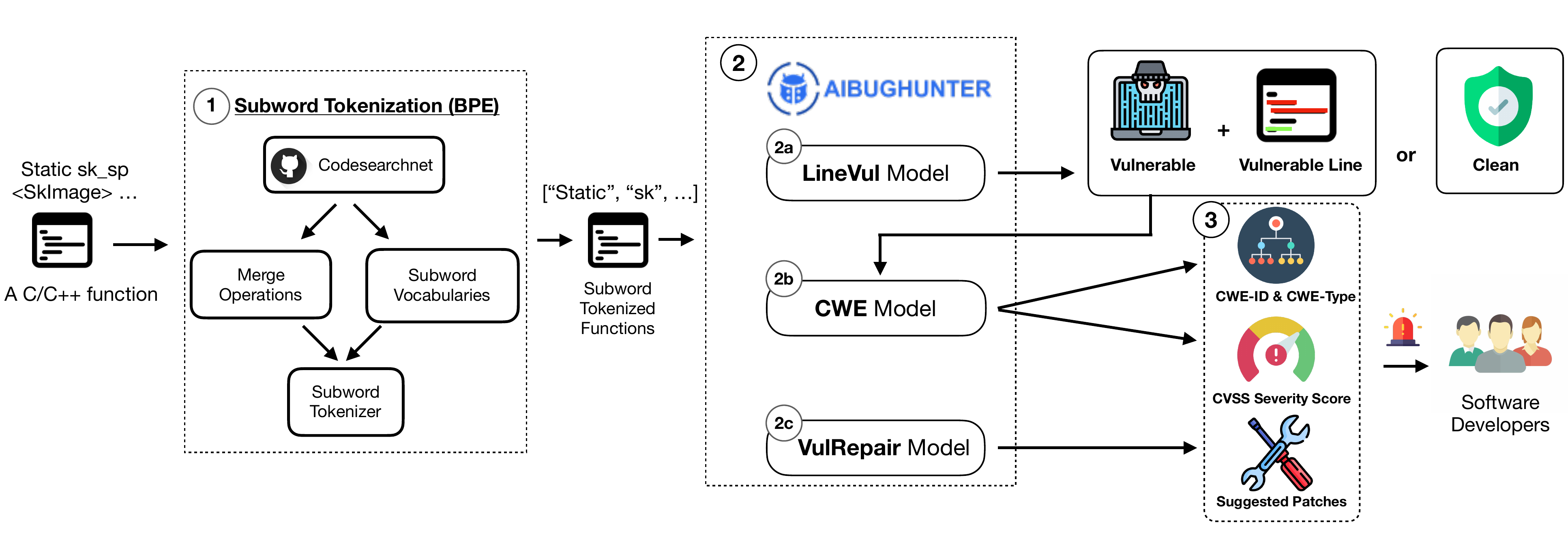}
\caption{An overview architecture of our approach.}
\label{fig:aibughunter}
\end{figure*}

\section{Learning to predict vulnerability type and severity}
\label{sec:our_approach}
Our approach is a vulnerability prediction framework consisting of three different inference tasks.
As shown in Fig~\ref{fig:aibughunter}, given a C/C++ function,
we first tokenize raw input into code tokens through Byte Pair Encoding (BPE) in Step \circled{1}.
In Step \circled{2}, the tokenized function is then input to a LineVul model proposed in our previous work~\citep{fu2022linevul} to predict vulnerable lines in the input function.
If vulnerable lines exist in the function,
our approach further predicts vulnerable types (i.e., CWE-ID and CWE-Type) and severity (i.e., CVSS severity score) of the vulnerable function as shown in step~\circled{2b}.
Furthermore, the vulnerable function is also input to the VulRepair~\citep{fu2022vulrepair} model to generate suggested repair patches as shown in step~\circled{2c}.
Finally in Step \circled{3}, \ourtool~integrates the predictions from LineVul, our approach, and VulRepair models and present them to software developers in the IDE.
We refer readers to our previous work~\citep{fu2022linevul} for more technical details about BPE tokenization and the Transformer architecture of our approach.

In this section, we introduce key new components in our \ourtool~approach over our prior works.
Given a vulnerable function, we aim to predict its vulnerability types, where CWE-ID and CWE-Type are available categorizations provided by~\cite{cwe}. CWE-Type is a higher-level of vulnerability category, where each CWE-Type may contain multiple similar CWE-IDs. Since CWE-ID and CWE-Type are highly correlated labels, we learn a shared CodeBERT model through multi-objective optimization as described in Section~\ref{moo_section}.

To predict the severity of vulnerabilities, we leverage a separate CodeBERT model instead of sharing the same model with the CWE classification task. This is due to (1) the CVSS severity score being a regression task that is different from the CWE classification; and (2) the CVSS severity score being determined using metrics provided by~\cite{cvss} rather than based on vulnerability types. Thus the vulnerability types and severity scores are not necessarily highly correlated.
In the following paragraphs, we describe in detail our approach for  CWE classification, followed by severity regression.

\begin{figure*}[t]
\includegraphics[width=\linewidth]{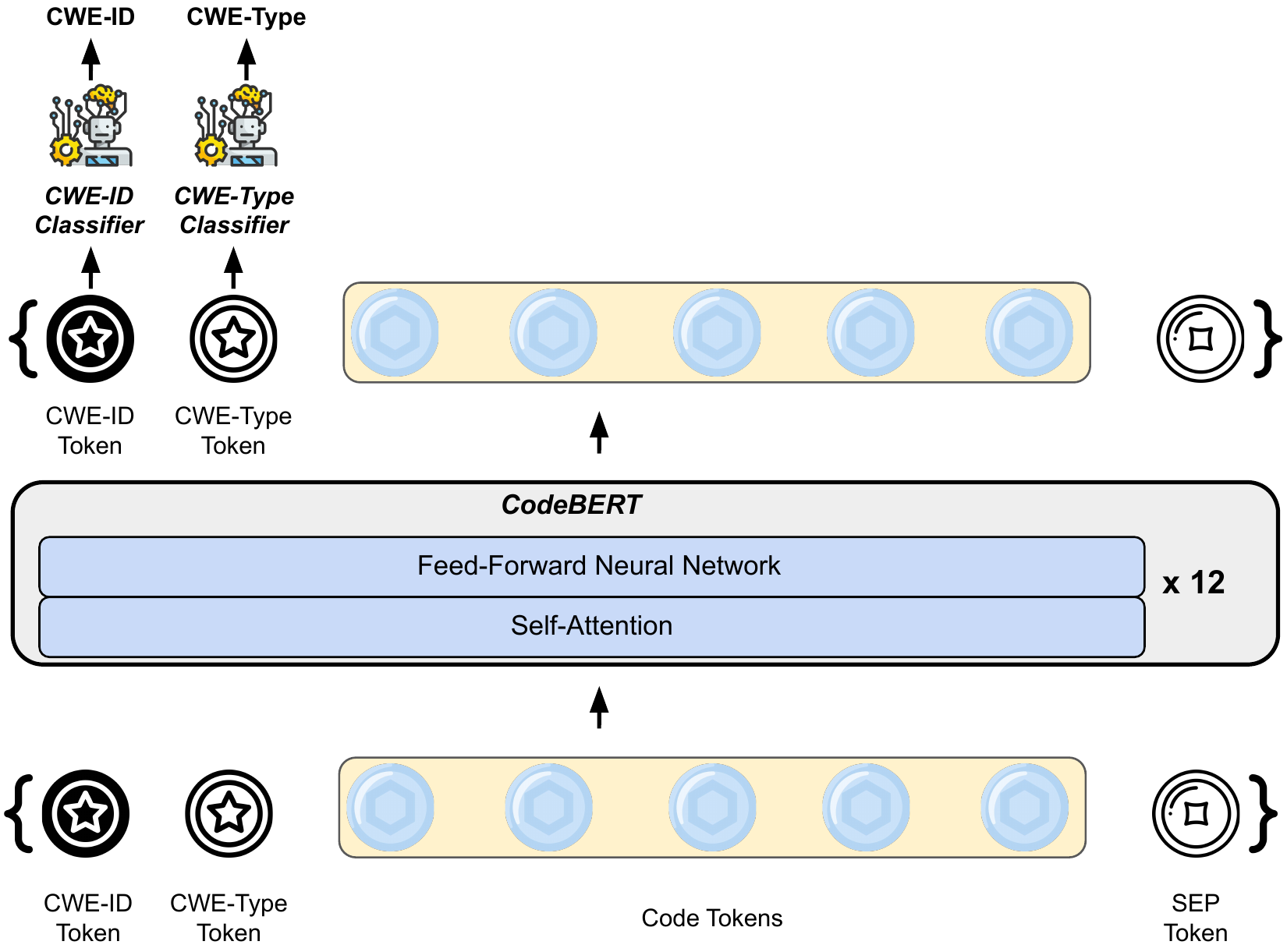}
\caption{An overview architecture of multi-objective CWE classification.}
\label{fig:movul}
\end{figure*}

\subsection{Multi-Objective CWE Classification}
\label{moo_section}
In this section, we introduce our novel multi-objective approach that is used to predict the CWE-ID and CWE-Type of a vulnerable function.

\subsubsection{Sequence Representation}
As shown in Fig~\ref{fig:movul}, instead of using only one ``[CLS]" token as a normal BERT model, our approach leverages two special tokens (one ``[CLS]" token for CWE-ID classification and the other ``[CLS\_TYPE]" token for CWE-Type classification) along with a ``[SEP]" token represents the end of a sequence. 
All of the special tokens are added during the subword tokenization process as described in our previous paper~\citep{fu2022linevul}.

The intuition behind using two special tokens for different tasks is the success of DeIT. ~\cite{touvron2021training} leveraged two special tokens to distill knowledge from a Transformer-based model for image classification tasks. 
In DeIT, one special token learns from the ground-truth labels while the other learns from the prediction generated by the teacher model to distill knowledge from it.
Similarly, our CWE-ID class token is responsible for the CWE-ID prediction and learns from ground-truth labels of CWE-ID while our CWE-Type class token focuses on CWE-Type prediction and learns from ground-truth labels of CWE-Type.

\subsubsection{Two Non-Shared Classification Heads}
Similar to DeIT~\citep{touvron2021training}, our approach uses two non-shared classification heads to generate predictions for two different tasks. Each classification head consists of two linear layers with dropout layers in between. Both heads rely on a softmax layer to activate the probabilities of each label which is the final prediction by our approach.
The parameters of the two heads are non-shared, so they are able to map the representation of their own special token (i.e., the class token of CWE-ID and CWE-Type) to the prediction without conflicting with each other.

The reasons for having two non-shared classification heads are (i) the number of classes for CWE-IDs is different from the number of classes for CWE-Type and (ii) we aim that each classification head can focus and specialize for each task (CWE-IDs or CWE-Type) to obtain better performances. Thus, we use separate non-shared heads to classify CWE-IDs and CWE-Type respectively.
In concurring with our design, the experiment results in Fig~\ref{fig:rq1_and_2} show that our multi-objective method with a shared transformer architecture achieves the best performance among other baseline methods.

\subsubsection{Multi-Objective Optimization}
\label{sec:movul_moo}

The problem solved by our approach can be considered as a multi-task learning (MTL) problem with an input space of $X$ and a collection of task spaces $\{y^{T}\}$ where $T$ is the number of tasks.
Specifically, we have a large vulnerability dataset with data points $\{x_{i}, y^{1}_{i}, y^{2}_{i}\}_i \in [N]$ where $x_{i}$ is a vulnerable function, $y^{1}$ is a CWE-ID label, $y^{2}$ is a CWE-Type label, and $N$ is the number of data points.

To optimize the parameters of a multi-task model, we need to minimize both loss functions yielded by CWE-ID and CWE-Type labels so the model can infer both labels given the same input.
Although the weighted summation is intuitively appealing as shown in Equation~\ref{equation:weightsum}, obtaining such weighted summation of loss functions for multi-task learning requires an expensive grid search over various scalings or the use of a heuristic such as~\cite{chen2018gradnorm, kendall2018multi} to find out the optimal values of $W_{1}$ and $W_{2}$. 

\begin{equation}
\mathcal{L}_{Total} = W_{1}\mathcal{L}_{ID} + W_{2}\mathcal{L}_{Type}
\label{equation:weightsum}
\end{equation}

Alternatively, our approach relies on the approach proposed by~\cite{sener2018multi} where the MTL problem is formulated as multi-objective optimization (MOO): optimizing a collection of possibly conflicting objectives. The training objective of our approach can be specified using a vector-valued loss L:

\begin{equation}
min \hspace{1.5mm} L(\theta^{sh},\theta^{1},\theta^{2}) = min \hspace{1.5mm} \big( \hat{\mathcal{L}}^{1}(\theta^{sh}, \theta^{1}), \hat{\mathcal{L}}^{2}(\theta^{sh}, \theta^{2})\big)
\label{equation:moo}
\end{equation}

where $L$ is the combined cross-entropy (CE) loss (described in Equation~\ref{equation:ce}) from both tasks computed by MOO, $\hat{\mathcal{L}}^{1}$ is the CE loss of the CWE-ID classification task, $\hat{\mathcal{L}}^{2}$ is the CE loss of the CWE-Type classification, $\theta^{sh}$ is parameters of shared 12-layer CodeBERT, $\theta^{1}$ is parameters of the CWE-ID classification head, and $\theta^{2}$ is parameters of the CWE-Type classification head as shown in Fig~\ref{fig:movul}.
In short, we aim to minimize all of the parameters (i.e., $\theta^{sh}, \theta^{1}, \theta^{2}$) during gradient descent simultaneously.

To fulfill the objective Equation~\ref{equation:moo} during the training phase of our approach, we leverage the same gradient update process as proposed by~\cite{sener2018multi}.
As shown in Algorithm~\ref{alg:mgda}, 
we first update the task-specific parameters (i.e., $\theta^{1} \hspace{1mm} and \hspace{1mm} \theta^{2}$) through the gradient descent algorithm. We then apply the Frank-Wolfe solver (please refer to the original paper written by~\cite{sener2018multi} for details) to find a common descent direction to satisfy our training objective. 
We then apply the solution of the Frank-Wolfe solver to update the shared parameters (i.e., $\theta^{sh}$) through the gradient descent algorithm. With such a gradient update process, all of the parameters (i.e., $\theta^{sh}$, $\theta^{1}, and \hspace{1mm}\theta^{2}$) can be updated at the same time without conflicting with each other. 


\begin{algorithm}
\caption{Gradient Update Equations for MTL}\label{alg:mgda}
\begin{algorithmic}[1]
\For{$t=1$ to T}
    \State $\theta^{t} = \theta^{t} - \eta \nabla_{\theta^{t}}\hat{\mathcal{L}}^{t}(\theta^{sh},\theta^{t})$ \Comment{Gradient descent on task-specific parameters(i.e., $\theta^{1}$, $\theta^{2}$)}
\EndFor
\State $\alpha^{1}, ..., \alpha^{T} = FRANKWOLFESOLVER(\theta)$ \Comment{Solve to find a common descent direction}
\State $\theta^{sh} = \theta^{sh} - \eta \sum_{t=1}^{T} \alpha^{t} \nabla_{\theta^{sh}}\hat{\mathcal{L}}^{t}(\theta^{sh},\theta^{t})$ \Comment{Gradient descent on shared parameters(i.e., $\theta^{sh}$)}
\end{algorithmic}
\end{algorithm}

\begin{figure*}[t]
\includegraphics[width=\linewidth]{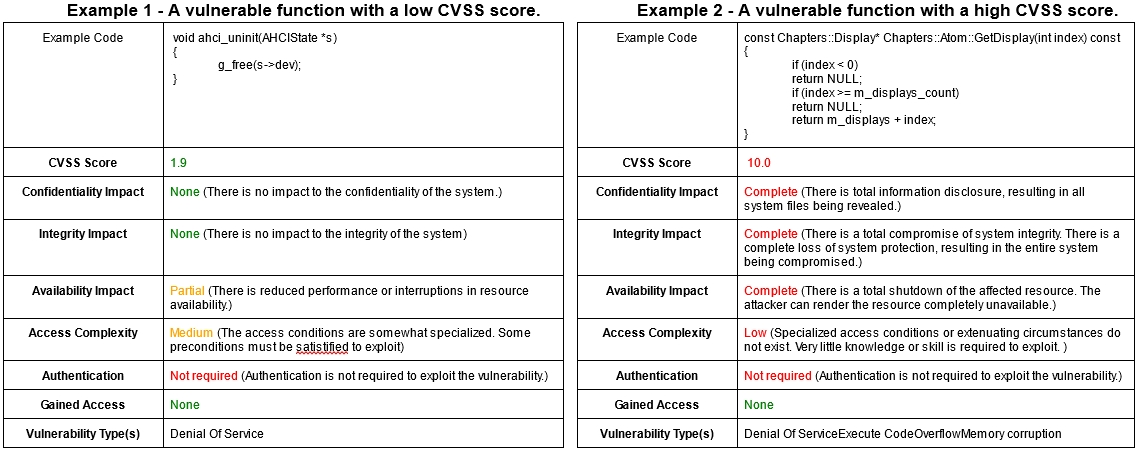}
\caption{Two concrete examples of high and low CVSS severity scores.}
\label{fig:cvss_examples}
\end{figure*}

\subsection{CVSS Severity Score Estimation}
We used Version 3.1 of the CVSS score which has a range of 0-10. Below, we provide two concrete examples and present the difference between high and low severity scores. It can be seen that the CVSS scores were assigned based on different measures such as confidentiality impact, integrity impact, availability impact, access complexity, authentication, and gained access etc. A low CVSS score (see Example 1 in Fig~\ref{fig:cvss_examples}) usually has None or Partial impact to the confidentiality, integrity, and availability aspects of the software system. In contrast, a high CVSS score (see Example 2 in Fig~\ref{fig:cvss_examples}) usually corresponds to higher impact such as Complete impact where there could be total information disclosure, total compromise of system integrity, and total shutdown of the affected resource.

As the pre-trained CodeBERT model has been demonstrated its effectiveness for vulnerability-related tasks~\citep{fu2022linevul, hin2022linevd}, we rely on CodeBERT to obtain word embeddings for each vulnerable function.
We add a linear layer as a regression head on top of CodeBERT, which returns one value for each vulnerable function as a severity score prediction.
We minimize the Mean Square Error (MSE) loss as described in Equation~\ref{equation:mse} to train the severity regression model:

\begin{equation}
\mathcal{L}_{MSE} = \frac{1}{n} \sum_{i=1}^{n}(y_{i}-\hat{y}_{i})^{2}
\label{equation:mse}
\end{equation}
where $y_{i}$ is a ground-truth severity score and $\hat{y}_{i}$ is a prediction of the model.
\section{A Quantitative Evaluation of \ourtool}
\label{sec:experiment}

In this section we present a quantitative evaluation of \ourtool.  We present our three research questions, our studied dataset, our experimental setup, and answers to our first three research questions along with their experimental results.

\subsection{Research Questions}

The empirical evaluation of LineVul and VulRepair backend components used in our \ourtool~have been presented in our previous works.
To evaluate our new proposed approach for vulnerability type and severity prediction,
we conduct a new set of experiments to compare our proposed method with existing baseline approaches.
Through an extensive evaluation of our approach on 8,783 C/C++ vulnerable functions including 88 different types of vulnerabilities, we answer the following three research questions:

\begin{enumerate}[{\bf RQ1:}]
\item {\bf \rqone}
We focus on  CWE-ID multi-class classification and compare our approach with four baseline models. Our approach achieves a multiclass accuracy of 0.65, which is 10\%-141\% more accurate than other baseline approaches with a median improvement of 86\%.
\item {\bf \rqtwo}
We focus on CWE-Type multiclass classification and compare our approach with the same four baseline models described in RQ1. Our approach achieves a multiclass accuracy of 0.74, which is 3\%-45\% more accurate than other baseline approaches with a median improvement of 23\%.
\item {\bf \rqthree}
We focus on the CVSS severity score regression task and compare our approach with 3 baseline approaches. Our approach achieves an MSE of 1.8479 and an MAE of 0.8753, which are better than the baseline approaches.
\end{enumerate}

\subsection{Studied Dataset}
To ensure a fair comparison with the previous work, we use the existing benchmark dataset~\citep{fan2020ac}.
We did not further parse data from 2020 to 2022 as previous studies did not publish scripts to collect datasets.
When implementing our data collection scripts, the collected data may not be the same as used by previous works, posing potential threats to internal validity.
Nevertheless, we encourage future studies to evaluate our approach on more recent datasets once available.

As this article is an extended version of our previous work~\citep{fu2022linevul}, we use the same experimental dataset (i.e., Big-Vul~\citep{fan2020ac}) to evaluate the performance of our approach on vulnerable functions.
The Big-Vul dataset is collected from 348 open-source Github projects, which includes 91 different CWEs from 2002 to 2019, and nearly 11k of C/C++ vulnerable functions.
Given a large number of vulnerable functions from diverse projects and timeframes, the Big-Vul dataset is a suitable dataset to evaluate whether our vulnerability classification and CVSS score estimation approaches can generalize well to the diverse samples.
Other vulnerability datasets such as the Devign dataset~\citep{zhou2019devign} are not selected because the CWE-ID and CVSS score information are not provided.

\subsection{Experimental Setup}
\noindent{\textbf{\underline{Data Splitting.}}} Similar to our previous work~\citep{fu2022linevul}, we split the dataset into 80\% of training data, 10\% of validation data, and 10\% of testing data.
We randomly split the data into three similar distributions so different vulnerability types are equally represented in training, validation, and testing sets. We also ensure that CWE-IDs appearing in the testing set should also appear in the training set.

\begin{table}[]
\caption{\revisedinline{Descriptive statistics of our studied datasets that describes the distribution of the severity score, and the distributions of cardinalities of CWE-ID and CWE-Type.}}
\centering
\label{tab:dataset_stat}
    \resizebox{\linewidth}{!}{\begin{tabular}{|c|ccccccc|}
\hline
                     & Mean & Median & Std. & 1st Quantile & 3rd Quantile & Min & Max  \\ \hline
CWE-ID Cardinality   & 100  & 9      & 281  & 3            & 49           & 1   & 2127 \\
CWE-Type Cardinality & 1255 & 415    & 1491 & 138          & 1827         & 1   & 4437 \\
Severity Score       & 6.18 & 6.8    & 1.95 & 4.6          & 7.5          & 1.2 & 10.0 \\ \hline
\end{tabular}}
\end{table}

\noindent{\textbf{\underline{Data Preprocessing.}}}
To satisfy the scenario of CWE classification tasks and the severity score regression task, we only keep the vulnerable functions with known CWE-ID, CWE-Type, and CVSS scores.
Table~\ref{tab:dataset_stat} presents the descriptive statistics of our studied dataset after removing non-vulnerable functions.
After data filtering, we keep 8,783 C/C++ functions with 88 different CWE-IDs, 6 different CWE-Types, and CVSS scores (labelled based on CVSS version 3.1~\citep{cvssv3}) ranging from 1.2-10.0.
Note that CWE-IDs and CWE-Types are many-to-one mappings where each CWE-ID has one CWE-Type but each CWE-Type may correspond to many CWE-IDs. 

\noindent{\textbf{\underline{Multi-objective Classification Model Implementation.}}} 
We leverage the pre-trained CodeBERT model as a backbone encoder to generate the shared representation of CWE-ID and CWE-Type classification tasks using the Transformers library in Python. 
We then add two classification heads on top of the backbone, one predicting the CWE-ID and the other predicting the CWE-Type.
Note that the parameters in the backbone are shared by both tasks, however, the parameters in each classification head are task-specific. 
We leverage two cross-entropy loss functions (i.e., $CE_{ID}$ and $CE_{Type}$) and implement the multi-objective optimization process based on the implementation provided by~\cite{sener2018multi} to fine-tune the CodeBERT model under the multi-task setting of CWE-ID and CWE-Type.
The multi-objective loss is implemented as described in Section~\ref{sec:movul_moo} where each cross-entropy loss is implemented based on Equation~\ref{equation:ce}. We use the PyTorch library to update the model and optimize the loss functions.

\begin{equation}
\mathcal{L}_{CE}(p,q) = -\sum_{x} p(x) \hspace{1mm} log_{q}(x)
\label{equation:ce}
\end{equation}

\noindent{\textbf{\underline{Severity Regression Model Implementation.}}} We leverage the pre-trained CodeBERT model with a regression head for CVSS score regression. The model is implemented with the Transformers library and trained using the PyTorch library. We use the Mean Squared Error (MSE) loss to update the model during training as Equation~\ref{equation:mse}.

\noindent{\textbf{\underline{Hyperparameter Settings for Fine-Tuning.}}}
We use the default setting of CodeBERT, i.e., 12 Transformer Encoder blocks, 768 hidden sizes, and 12 attention heads. 
We follow the same fine-tuning strategy provided by~\cite{feng2020codebert}. During training, the learning rate is set to 2e-5 with a constant schedule. 
We use backpropagation with AdamW optimizer~\citep{loshchilov2017decoupled} which is widely adopted to fine-tune Transformer-based models to update the model and minimize the loss function.
The best model is selected based on the validation data, which will perform inference on testing data as final evaluation results.

\noindent{\textbf{\underline{Execution Environment.}}}
All of the experiments are run under Ubuntu 20.04 system with an AMD Ryzen 9 5950X CPU with 16C/32T, 64GB RAM, and an NVIDIA GTX 3090 GPU (24GB of memory).

\subsection{Experimental Results}

\begin{figure}[t]
\includegraphics[width=\linewidth]{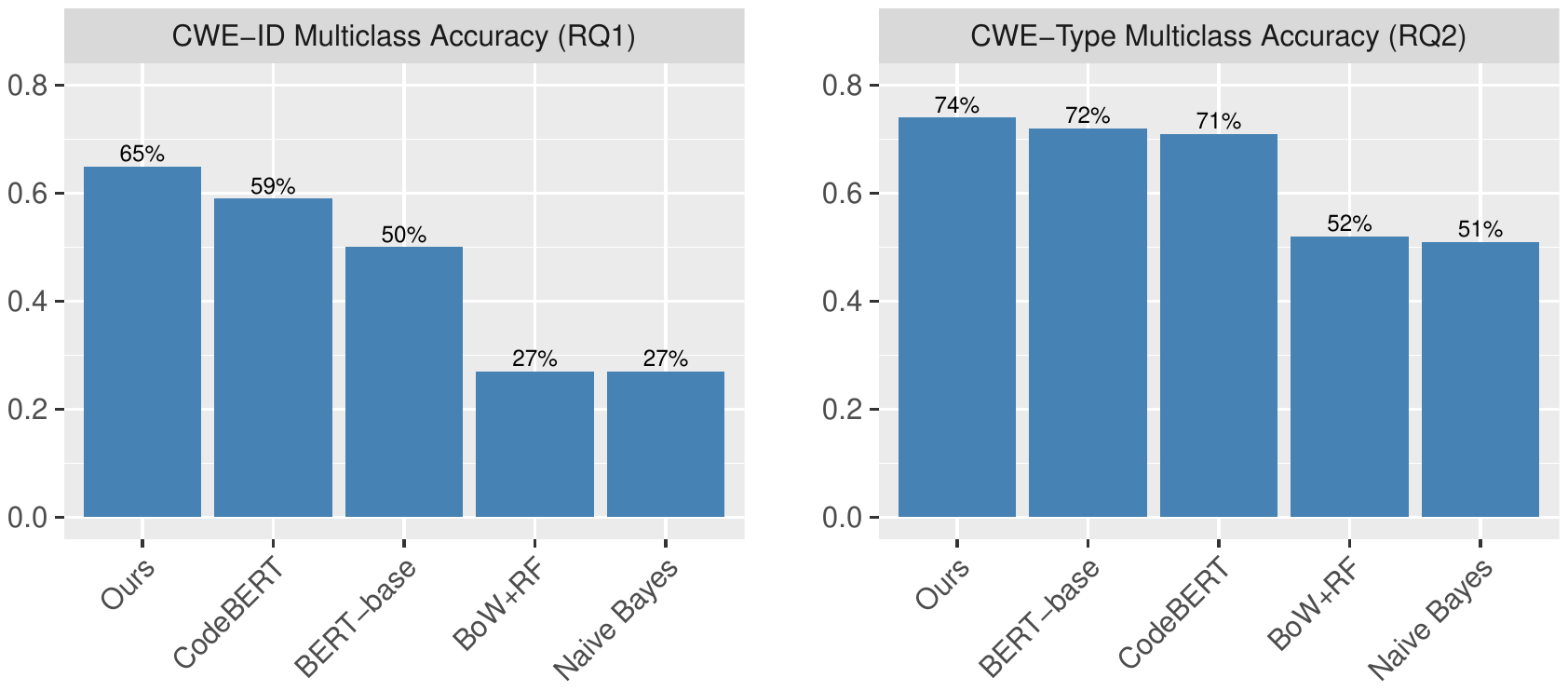}
\caption{(RQ1 and RQ2) The Multiclass Accuracy of our approach and four other baselines. ($\nearrow$) Higher Multiclass Accuracy = Better.}
\label{fig:rq1_and_2}
\end{figure}

\subsection*{\textbf{RQ1: \rqone}}
\noindent{\textbf{\underline{Approach.}}} To answer this RQ, we focus on  CWE-ID multi-class classification and compare our approach with four baseline models, described as follows: 
\begin{enumerate}
    \item BERT models pre-trained on natural language (i.e., BERT-base~\citep{devlin2018bert}), which have been adopted for CWE classification tasks~\citep{das2021v2w, wang2021towards}. 
    \item BERT models pre-trained on programming language (i.e., CodeBERT~\citep{feng2020codebert}), which have been applied to software vulnerability prediction~\citep{fu2022linevul, thapa2022transformer, yuan2022deep}.
    \item BoW+RF uses bag of words as features together with a Random Forest model for CWE-ID classification~\citep{aota2020automation, wang2020machine}.
    \item BoW+NB uses bag of words as features together with a Naive Bayes model for CWE-ID classification~\citep{na2016study}.
\end{enumerate}
The pre-trained BERT-based language models are selected because previous studies such as~\cite{wang2021towards} and~\cite{zhu2022bert} have used them to achieve promising results on the CWE-ID classification tasks. The Random Forest and Naive Bayes models are selected because they are important machine learning-based methods for CWE-ID classification tasks proposed in previous studies.

We evaluate our approach based on the multiclass accuracy which is computed as $\frac{\textrm{Correctly Predicted Testing Data}}{\textrm{Total Testing Data}}$.

\vspace{1mm}

\noindent{\textbf{\underline{Results.}}} Fig~\ref{fig:rq1_and_2} presents the experimental results of our approach and the four baseline approaches according to the multiclass accuracy. 

\textbf{Our approach achieves an accuracy of 0.65, which is 10\%-141\% more accurate than other baseline approaches with a median improvement of 86\%.}
These results confirm that our approach is more accurate than other baseline approaches for CWE-ID classification.

We use CodeBERT as our backbone architecture, however, our approach outperforms the CodeBERT model by 6\%.
Our approach can learn knowledge from two perspectives based on the class of CWE-ID and the class of CWE-Type where both classes describe the same vulnerable function.
The correlated information between the two kinds of labels further benefits our method.
On the other hand, the CodeBERT method only learns from CWE-ID labels.
In other words, the comparison between our approach and CodeBERT highlights the advancement of using labels from both tasks (i.e., CWE-ID and CWE-Type) with multi-objective optimization.
In short, our results demonstrate that \textbf{the multi-task learning with multi-objective optimization using both CWE-ID and CWE-Type labels outperforms other baselines that are only trained using CWE-ID label.}

We analyze the performance of our approach on 879 testing samples. First, 567 of 879 (65\%) are correctly predicted. On the other hand, 312 samples are misclassified. Among the 312 misclassified samples, we find that 89 of 312 (35\%) were predicted as close to the ground truth (i.e., incorrectly predict the CWE-ID, but correctly predict the CWE-Type).
This means our approach can at least correctly predict the vulnerability type for 75\% $\frac{(567+89)}{879}$ of testing samples (outperform all other baselines), highlighting the potential usefulness of our approach in practice.

\begin{table}[t]
 \caption{(RQ1 Discussion) The Accuracy of our approach~for the Top-25 Most Dangerous CWEs (\url{https://cwe.mitre.org/top25/archive/2021/2021_cwe_top25.html}).}
\label{tab:movul_cwe_list}

\centering

\begin{tabular}{ccccc}
\hline
\multicolumn{1}{|c|}{Rank} & \multicolumn{1}{c|}{CWE-ID}  & \multicolumn{1}{c|}{Name}                            & \multicolumn{1}{c|}{Accuracy} & \multicolumn{1}{c|}{Proportion} \\ \hline
\multicolumn{1}{|c|}{1}    & \multicolumn{1}{c|}{CWE-787} & \multicolumn{1}{c|}{Out-of-bounds Write}             & \multicolumn{1}{c|}{43\%}     & \multicolumn{1}{c|}{9/21}       \\ \hline
\multicolumn{1}{|c|}{2}    & \multicolumn{1}{c|}{CWE-79}  & \multicolumn{1}{c|}{Cross-site Scripting}            & \multicolumn{1}{c|}{29\%}     & \multicolumn{1}{c|}{2/7}        \\ \hline
\multicolumn{1}{|c|}{3}    & \multicolumn{1}{c|}{CWE-125} & \multicolumn{1}{c|}{Out-of-bounds Read}              & \multicolumn{1}{c|}{67\%}     & \multicolumn{1}{c|}{44/66}      \\ \hline
\multicolumn{1}{|c|}{4}    & \multicolumn{1}{c|}{CWE-20}  & \multicolumn{1}{c|}{Improper Input Validation}       & \multicolumn{1}{c|}{66\%}     & \multicolumn{1}{c|}{71/107}     \\ \hline
\multicolumn{1}{|c|}{7}    & \multicolumn{1}{c|}{CWE-416} & \multicolumn{1}{c|}{Use After Free}                  & \multicolumn{1}{c|}{52\%}     & \multicolumn{1}{c|}{15/29}      \\ \hline
\multicolumn{1}{|c|}{8}    & \multicolumn{1}{c|}{CWE-22}  & \multicolumn{1}{c|}{Path Traversal}                  & \multicolumn{1}{c|}{0\%}      & \multicolumn{1}{c|}{0/4}        \\ \hline
\multicolumn{1}{|c|}{9}    & \multicolumn{1}{c|}{CWE-352} & \multicolumn{1}{c|}{Cross-Site Request Forgery}      & \multicolumn{1}{c|}{0\%}      & \multicolumn{1}{c|}{0/1}        \\ \hline
\multicolumn{1}{|c|}{12}   & \multicolumn{1}{c|}{CWE-190} & \multicolumn{1}{c|}{Integer Overflow}                & \multicolumn{1}{c|}{68\%}     & \multicolumn{1}{c|}{21/31}      \\ \hline
\multicolumn{1}{|c|}{14}   & \multicolumn{1}{c|}{CWE-287} & \multicolumn{1}{c|}{Improper Authentication}         & \multicolumn{1}{c|}{0\%}      & \multicolumn{1}{c|}{0/2}        \\ \hline
\multicolumn{1}{|c|}{15}   & \multicolumn{1}{c|}{CWE-476} & \multicolumn{1}{c|}{NULL Pointer Dereference}        & \multicolumn{1}{c|}{41\%}     & \multicolumn{1}{c|}{7/17}       \\ \hline
\multicolumn{1}{|c|}{17}   & \multicolumn{1}{c|}{CWE-119} & \multicolumn{1}{c|}{Improper Restriction}            & \multicolumn{1}{c|}{79\%}     & \multicolumn{1}{c|}{180/228}    \\ \hline
\multicolumn{1}{|c|}{18}   & \multicolumn{1}{c|}{CWE-862} & \multicolumn{1}{c|}{Missing Authorization}           & \multicolumn{1}{c|}{0\%}      & \multicolumn{1}{c|}{0/1}        \\ \hline
\multicolumn{1}{|c|}{20}   & \multicolumn{1}{c|}{CWE-200} & \multicolumn{1}{c|}{Exposure of Sensitive Info}      & \multicolumn{1}{c|}{62\%}     & \multicolumn{1}{c|}{26/42}      \\ \hline
\multicolumn{1}{|c|}{22}   & \multicolumn{1}{c|}{CWE-732} & \multicolumn{1}{c|}{Incorrect Permission Assignment} & \multicolumn{1}{c|}{86\%}     & \multicolumn{1}{c|}{6/7}        \\ \hline
\multicolumn{1}{|c|}{23}   & \multicolumn{1}{c|}{CWE-611} & \multicolumn{1}{c|}{Improper Restriction}            & \multicolumn{1}{c|}{50\%}     & \multicolumn{1}{c|}{1/2}        \\ \hline
\multicolumn{1}{|c|}{25}   & \multicolumn{1}{c|}{CWE-77}  & \multicolumn{1}{c|}{Improper Neutralization}         & \multicolumn{1}{c|}{0\%}      & \multicolumn{1}{c|}{0/1}        \\ \hline
\multicolumn{1}{|c|}{}     & \multicolumn{1}{c|}{}        & \multicolumn{1}{c|}{}                                & \multicolumn{1}{c|}{67\%}     & \multicolumn{1}{c|}{382/566}    \\ \hline
\multicolumn{1}{l}{}       & \multicolumn{1}{l}{}         & \multicolumn{1}{l}{}                                 & \multicolumn{1}{l}{}          & \multicolumn{1}{l}{}           
\end{tabular}
\end{table}
\begin{figure}[t]
\includegraphics[width=\linewidth]{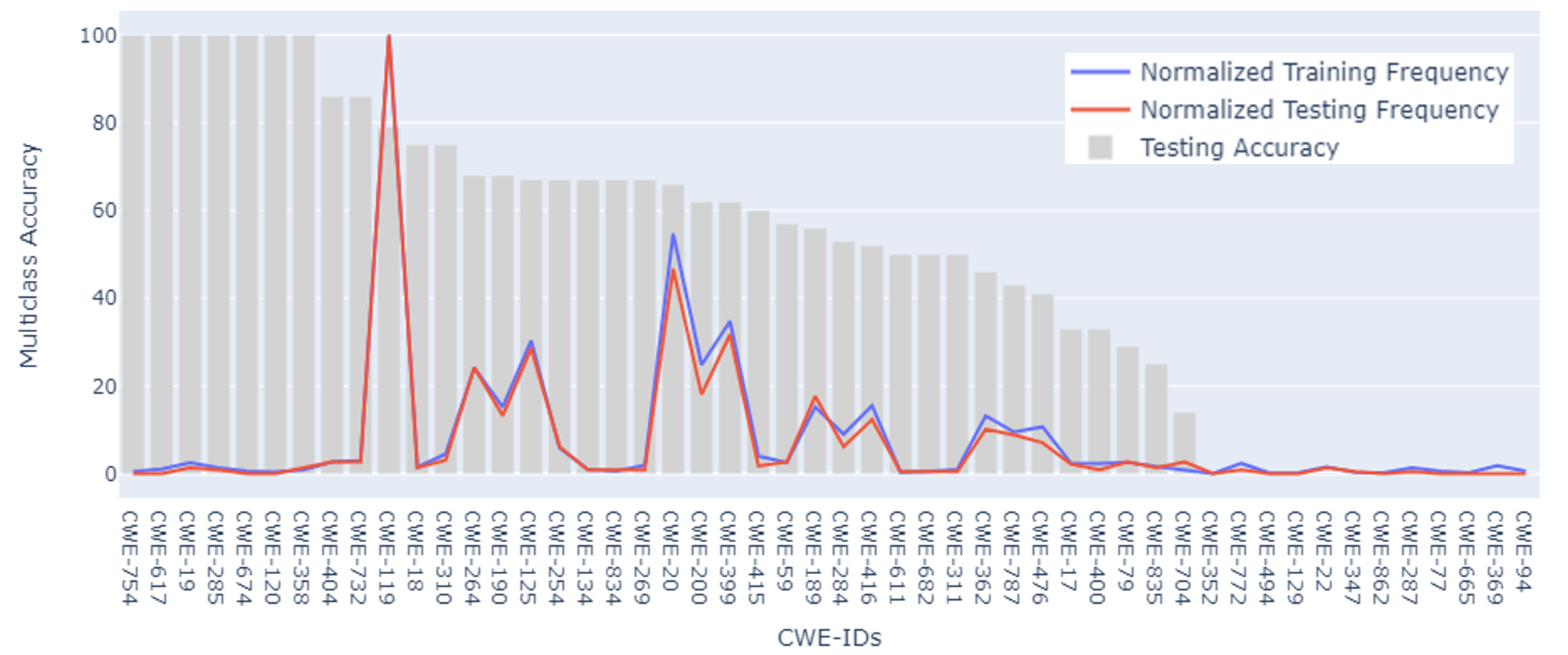}
\caption{(RQ1 Discussion) Our method's Multiclass Accuracy of CWE-ID classification for each CWE-ID in the testing set. The accuracy is shown in percentage.}
\label{fig:cwe_id_discussion}
\end{figure}

To further investigate whether our approach can classify dangerous real-world vulnerabilities,
we further evaluate our approach on Top-25 most dangerous CWE-IDs~\citep{cwetop25} in the testing set to understand the significance of our approach in the practical usage scenario.
Table~\ref{tab:movul_cwe_list} presents the accuracy of our approach on Top-25 most dangerous CWE-IDs.
\textbf{We find that our approach can correctly predict 67\% of the vulnerable functions affected
by the Top-25 most dangerous CWE-IDs, which is better than the average performance of our approach(i.e., 65\%).}

In addition, Fig~\ref{fig:cwe_id_discussion} presents our method's accuracy for each CWE-ID in the testing set. It can be seen that the accuracy of our approach is not highly correlated to training or testing data frequencies. Our approach performs well on some of the CWE-IDs with low frequencies such as CWE-754 while having challenges generalizing to other low frequencies CWE-IDs such as CWE-94. However, those CWE-IDs that cannot be identified by our approaches are all CWE-IDs rarely occur in the dataset. This highlights the challenge of imbalanced data in the CWE-ID classification task where some CWE-IDs are common (e.g., CWE-119) and easy to collect while other CWE-IDs can be rare (e.g., CWE-369) and difficult to collect. Those rare CWE-IDs are more prone than common CWE-IDs to be misclassified by our approach due to not-enough training samples. Thus, future researchers may explore new techniques to solve this imbalance problem.

Last but not least, we found that the complexity of vulnerabilities may also affect the performance of our approach.
In particular, our approach achieves an accuracy of 86\% for the least complex CWE-IDs that are under the CWE-Type of the ``class weakness''.
Class weaknesses typically describe issues in terms of 1 or 2 of the following dimensions: behavior, property, and resource~\citep{classweakness}.
For instance, the class weakness of "Uncontrolled Resource Consumption" (CWE-400) describes an issue (Uncontrolled) with a behavior (Consumption) associated with any type of resource.
However, our approach only achieves an accuracy of 51\% for the most complex CWE-IDs that are under the CWE-Type of the ``variant weakness''.
Variant weaknesses typically describe issues in terms of 3 to 5 of the following dimensions: behavior, property, technology, language, and resource~\citep{variantweakness}.
For instance, the variant weakness of "Use After Free" (CWE-416) describes an issue (Referencing memory after it has been freed) with a specific resource (Memory) with specific languages (C/C++).
These results highlight the challenge of classifying those complex vulnerability types such as the variant weakness.

\begin{figure}[t]
\includegraphics[width=\linewidth]{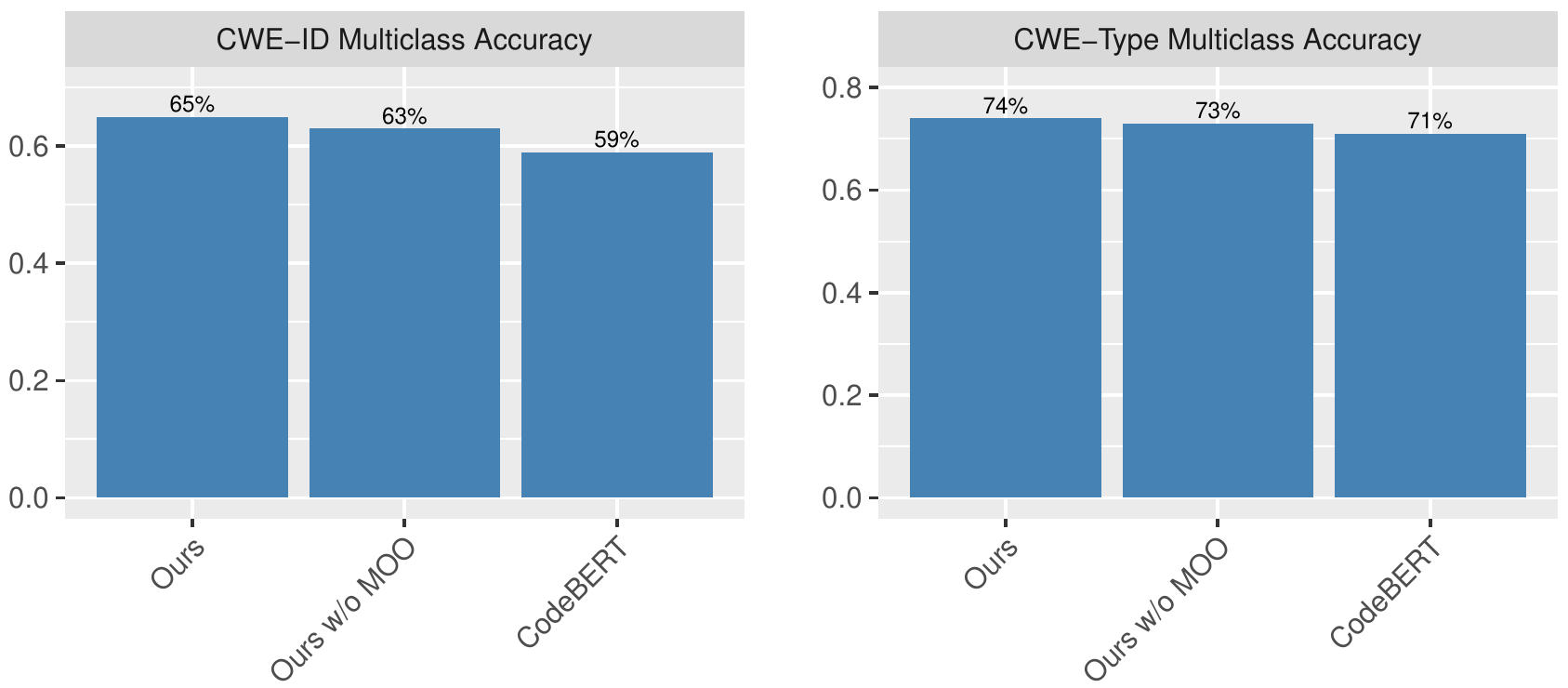}
\caption{The Multiclass Accuracy of our approach, our approach w/o MOO, and single-task CodeBERT. ($\nearrow$) Higher Multiclass Accuracy = Better.}
\label{fig:discussion}
\end{figure}

\begin{table}[]
\caption{\revisedinline{(RQ2) The Multiclass Accuracy of each CWE type for all of the approaches evaluated in RQ2.}}
\centering
\label{tab:cwe_type_performance}
    \resizebox{\linewidth}{!}{
    \begin{tabular}{|c|cccccc|c|}
    \hline
    Methods     & \multicolumn{6}{c|}{CWE Abstract Types}                                                                      &               \\ \hline
                & Class            & Category         & Variant          & Base             & Pillar           & Deprecated    & Overall       \\ \hline
    Ours        & 85.55\%          & 67.01\%          & \textbf{62.86\%} & 60.8\%           & 52.94\%          & \textbf{50\%} & \textbf{74\%} \\ \hline
    BERT-base   & 84.42\%          & \textbf{68.53\%} & 34.29\%          & 55.11\%          & \textbf{64.71\%} & \textbf{50\%} & 72\%          \\ \hline
    CodeBERT    & 78.78\%          & 63.95\%          & 42.86\%          & \textbf{71.02\%} & 52.94\%          & 30\%          & 71\%          \\ \hline
    BoW+RF      & 74.49\%          & 29.44\%          & 14.29\%          & 33.52\%          & 5.88\%           & 10\%          & 52\%          \\ \hline
    Naive Bayes & \textbf{99.77\%} & 3.55\%           & -                & -                & -                & -             & 51\%          \\ \hline
    \end{tabular}}
\end{table}
\begin{table}[]
\caption{\revisedinline{(RQ2 Discussion) The analysis of how different function lengths affect the multi-class accuracy of our approach for CWE-ID and CWE-Type prediction tasks. Note. Function lengths counted by number of tokens in a tokenized function.}}
\centering
\label{tab:func_length}
\begin{tabular}{|c|c|c|}
\hline
Function Length (Tokens) & CWE-ID Accuracy & CWE-Type Accuracy \\ \hline
0-100                    & 84\%            & 85\%              \\ \hline
101-200                  & 72\%            & 81\%              \\ \hline
201-300                  & 69\%            & 71\%              \\ \hline
301-400                  & 60\%            & 69\%              \\ \hline
401-500                  & 61\%            & 70\%              \\ \hline
\textgreater{}500        & 57\%            & 72\%              \\ \hline
\end{tabular}
\end{table}

\subsection*{\textbf{RQ2: \rqtwo}}
\noindent{\textbf{\underline{Approach.}}} To answer this RQ, we focus on CWE-Type multiclass classification and compare our approach with the same four baseline models described in RQ1. We adopt the same measure as mentioned in RQ1 to evaluate our approach.

\noindent{\textbf{\underline{Results.}}} Fig~\ref{fig:rq1_and_2} presents the experimental results of our approach and the four baseline approaches according to the multiclass accuracy.

\textbf{Our approach achieves an accuracy of 0.74, which is 3\%-45\% more accurate than other baseline approaches with a median improvement of 23\%.}
These results confirm that our approach is more accurate than other baseline approaches for CWE-Type classification.

Our approach performs the best, which is the only model that leverages both CWE-Type and CWE-ID labels during training.
The improvement of our approach compared with other baseline approaches is 3\%-45\% which is not as significant as the improvement demonstrated in RQ1 (10\%-141\%).
The difference in improvements implies that while leveraging both labels can benefit performance for both CWE-ID and CWE-Type classification tasks, our method is more beneficial for the CWE-ID classification.

In addition, Table~\ref{tab:cwe_type_performance} presents detailed accuracy for each CWE-Type.
It can be seen that the performance depends on the number of samples and varies for each type.
Nevertheless, our approach has the best overall accuracy and is the only approach that achieves at least 50\% of accuracy for each CWE-Type.


In Section~\ref{sec:movul_moo}, we proposed leveraging Multi-Objective Optimization (MOO) instead of taking the weighted summation of loss functions for gradient descent.
We now further evaluate whether MOO can help our approach learn better on multi-task learning. 
Specifically, we compare our approach (with using MOO) with a variant method (without using MOO) that leverages a weighted summary of the loss function during gradient descent.
The loss function of the weighted summary version of our approach is described as Equation~\ref{equation:weightsum}.
We set $W1$ and $W2$ to 0.5 so both tasks contribute equally to the total loss.
To ensure a fair comparison, we only switch the MOO component of our approach and adopt the same model architecture, hyperparameters, and training strategy for both approaches.

Fig~\ref{fig:discussion} presents the accuracy of our approach, the variant approach, and the single-task CodeBERT.
\textbf{We find that the multi-task learning framework is always better than the CodeBERT which only learns from a single task, and our approach performs the best on both tasks.}
Our approach can achieve an accuracy of 63\%-65\% and 73\%-74\% on CWE-ID and CWE-Type classification respectively while single-task CodeBERT only achieves an accuracy of 59\% and 71\%.
This result confirms that (1) leveraging multi-task learning on two correlated tasks may benefit the model performance on both tasks and (2) the MOO approach used by our approach can learn a model with higher accuracy than the weighted summary approach.

Furthermore, we analyze the impact of function length on our approach for CWE-ID and CWE-Type classification. According to Table~\ref{tab:func_length}, when the function length is short, e.g., consisting of 0-100 tokens, our tool can have better 84\% and 85\% accuracy respectively. However, the performance decreases as functions become longer, for functions consisting of more than 500 tokens, the accuracy becomes 57\% and 72\% respectively. These results highlight the challenge of tackling long sequences for vulnerability classification tasks. Thus, future researchers should further explore techniques that can classify difficult longer vulnerable functions.

\subsection*{\textbf{RQ3: \rqthree}}
\noindent{\textbf{\underline{Approach.}}} To answer this RQ, we focus on the CVSS severity score regression task and compare our approach with 3 baseline approaches as follows:
\begin{enumerate}
    \item BERT models pre-trained on natural language (i.e., BERT-base~\citep{devlin2018bert}). 
    \item BoW+RF uses bag of words as features together with a Random Forest model for severity score regression~\citep{aota2020automation, wang2020machine}.
    \item BoW+LR uses bag of words as features together with a Linear Regression model for severity score regression.
\end{enumerate}
We evaluate our approach based on Mean Squared Error (MSE)~\ref{equation:mse} and Mean Absolute Error (MAE) where MSE penalizes the predictions that are far from true values through the square of Euclidean distance and MAE measures the exact distance between predicted values and ground-truth values as $MAE = \frac{1}{n}\sum_{i=1}^{n}\mid y_{i} - \hat{y}_{i} \mid$.

\begin{figure}[t]
\includegraphics[width=\linewidth]{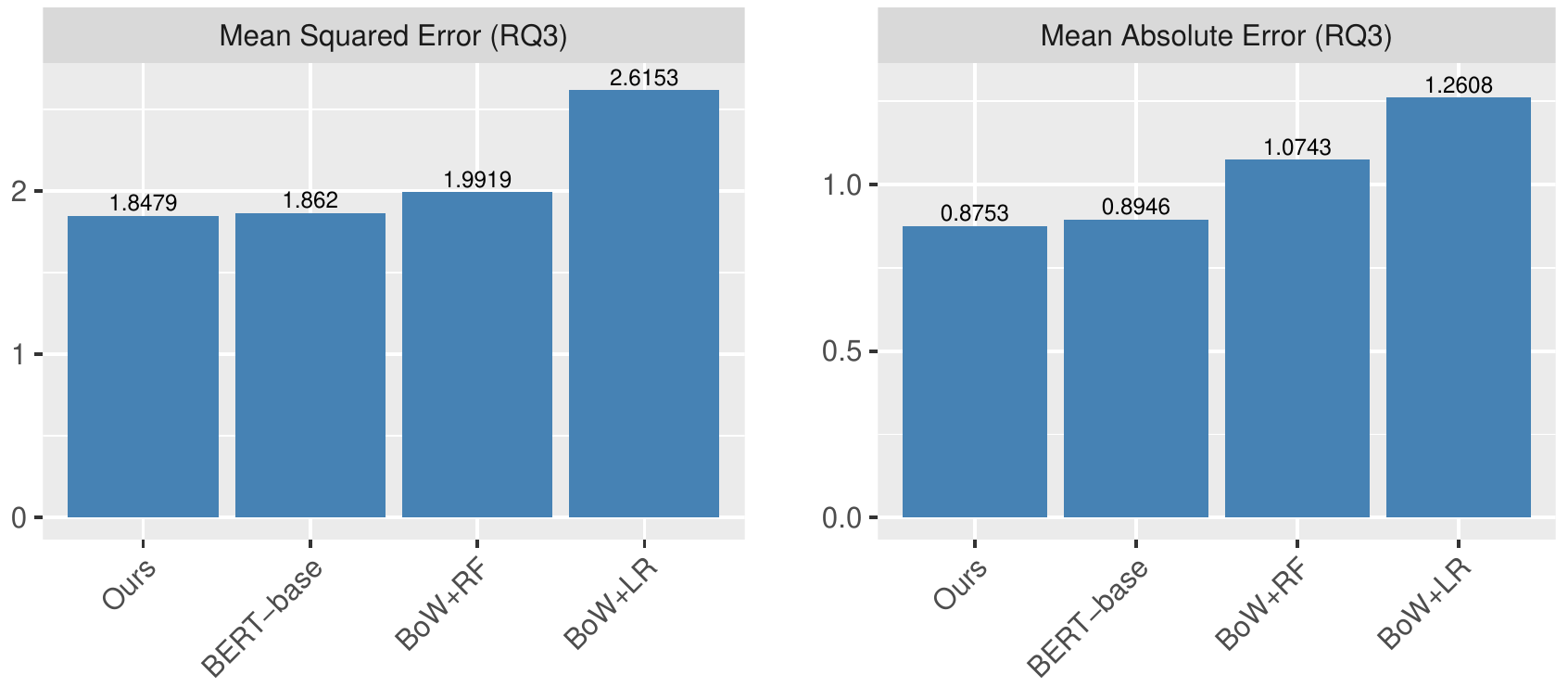}
\caption{(RQ3) The Mean Squared Error (MSE) and Mean Absolute Error (MAE) of our approach and three other baselines. ($\searrow$) Lower MSE, MAE = Better.}
\label{fig:rq3}
\end{figure}

\begin{figure}[t]
\includegraphics[width=\linewidth]{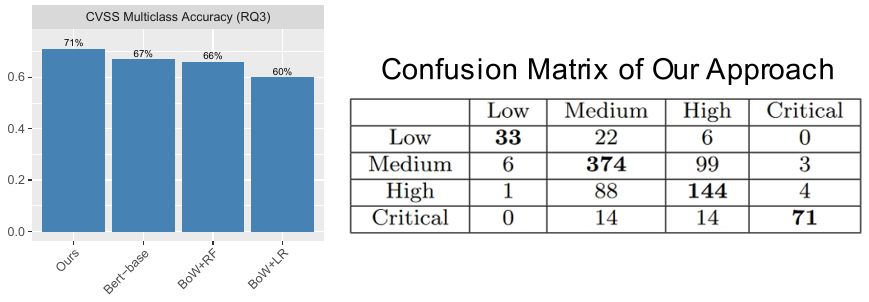}
\caption{(RQ3 Discussion) The left part is the multi-class accuracy of the CVSS score for each approach evaluated in RQ3. The right part is the confusion matrix of our approach. Note that each class of CVSS is directly mapped from the CVSS score as shown at the bottom of the confusion matrix table.}
\label{fig:cvss_cls}
\end{figure}

\noindent{\textbf{\underline{Results.}}} Fig~\ref{fig:rq3} presents the experimental results of our approach and the three baseline approaches according to the MSE and MAE loss.

\textbf{Our approach achieves an MSE of 1.8479 and an MAE of 0.8753, which are better than all of the baseline approaches.}
These results confirm that our approach can predict the most accurate severity scores.

Our approach \textbf{achieves 0.1440 and 0.7674 less MSE than the baselines using traditional Bag of Words and Machine Learning algorithms (i.e., BoW+RF, BoW+LR)}.
This result highlights the advancement of leveraging a BPE tokenization and pre-trained word embedding with a Transformer-based architecture.
The word embeddings with the self-attention mechanism~\citep{vaswani2017attention} in the Transformer model can learn the semantic features of input source code while the traditional BoW approach only considers the word frequencies when representing source code. Thus, our approach learns a more accurate mapping between a vulnerable function and its corresponding severity score.

Our approach \textbf{achieves 0.0141 less MSE and 0.0193 less MAE than the BERT-base (pre-trained on natural language) model}.
This result confirms that leveraging a BERT architecture pre-trained using programming languages (our approach) can improve the one pre-trained using natural language.

To investigate whether each approach can accurately predict the severity of vulnerable functions, we map the CVSS score into four classes of severity based on the CVSS protocol, i.e., low, medium, high, and critical as detailed in Fig~\ref{fig:cvss_cls}.
Our approach achieves an accuracy of 0.71, which is 6\%-18\% better than other baselines.
The result confirms that our approach can correctly predict the severity class for 71\% of vulnerable functions in testing data.

To further investigate our approach's performance, we present our approach's confusion matrix in Fig~\ref{fig:cvss_cls}.
It can be seen that our approach neither estimates low severity as critical (last row, first column) nor estimates critical severity as low (first row, last column).
Furthermore, the last column shows that when our approach predicts a critical severity, the accuracy is 91\%.
Nevertheless, the most common error of our approach is predicting samples to the close class such as estimating a medium severity as high severity, which highlights the challenge of the CVSS severity estimation task.
\section{Qualitative Evaluations of \ourtool}
\label{sec:survey}

We conducted qualitative evaluations including (1) a survey study to obtain software practitioners' perceptions of our \ourtool~tool; and (2) a user study to investigate the impact that our \ourtool~could have on developers’ productivity in security aspects, to answer the following research question:

\begin{enumerate}[{\bf RQ4: }]
\item {\bf \rqfour}
According to our survey study, each kind of vulnerability prediction provided by our \ourtool~is perceived as useful by 47\%-86\% of participated software practitioners. Furthermore, 90\% of participants consider adopting our \ourtool~if it is freely available in an IDE without conditions.
Moreover, our user study shows that our \ourtool~could save developers’ time spent on security analysis that could enhance security productivity during software development.
\end{enumerate}

\subsection{A Qualitative Survey Study}
Following ~\cite{kitchenham2008personal}, we conduct our study according to the following steps: (1) design and develop a survey, (2) recruit and select participants, and (3) verify data and analyze data. We explain the detail of each step below.

\subsubsection{Survey Design}
\textbf{Step 1: Design and development of the survey: }
We design our survey as a cross-sectional study where participants provided their responses at one fixed point in time.
The survey consists of 6 closed-ended questions and 5 open-ended questions.
For closed-ended questions, we use multiple-choice questions and a Likert scale from 1 to 5.
Our survey consists of two parts: preliminary questions and developers' perceptions of AI-based software vulnerability predictions.

\textit{Part I: Demographics.} The survey starts with a question, (``(D1) What is your role in your software development team?''), to ensure that our survey results are obtained from the right target participants.
Then, the survey is followed by a demographics question, (``(D2) What is the level of your professional experience?''), to ensure our survey is distributed across software practitioners with different levels of professional experience.

\textit{Part II: Vulnerability predictions generated by our \ourtool.} We then ask about software practitioners' perceptions of AI-based vulnerability predictions.
Specifically, we present an example visualization of a prediction generated by \ourtool~as shown in Figure~\ref{fig:aib_ui}.
Then, we ask four questions, i.e.,
 (``(Q1) How do you perceive the usefulness of the recommended location of the vulnerability (i.e., line number)?''),
 (``(Q2) How do you perceive the usefulness of the vulnerability severity prediction?''),
 (``(Q3) How do you perceive the usefulness of the vulnerability type prediction (i.e., CWE-ID and CWE-Type)?''),
 and (``(Q4) How do you perceive the usefulness of the ``Quick Fix'' button which will replace a vulnerable line with the suggested repair on click?'')
Each question is followed by an open question for the rationale.

We use Google Form to conduct our survey in an online setting.
Each participant is provided with an explanatory statement on the landing page that describes the purpose of the study, why the participant is chosen for this study, possible benefits and risks, and confidentiality. 
The survey takes approximately 10 minutes to complete and is completely anonymous.
Our survey has been rigorously reviewed and approved by the Monash University Human Research Ethics Committee (MUHREC ID: 35047).

\textbf{Step 2: Recruit and select participants: }
We recruit developers that have software development experience through LinkedIn and Facebook platforms.
We send a survey invitation to the target groups via direct message. 
To mitigate potential bias introduced by the participant groups, we selected participants with different software engineering-related professions, different lengths of professional experience, and different organizations.
Finally, we obtained a total of 22 responses over a two-week period of recruitment.

\textbf{Step 3: Verify data and analyze data: }
To verify the completeness of the response in our survey (i.e., whether all questions were appropriately answered), we manually review all of the open-ended questions. 
Finally, we obtain a set of 21 valid responses.
We present the results of closed-ended responses in a Likert scale with stacked bar plots. 
We manually analyze the responses to the open-ended questions to better understand the in-depth insights.

\begin{figure*}[t]
\includegraphics[width=\linewidth]{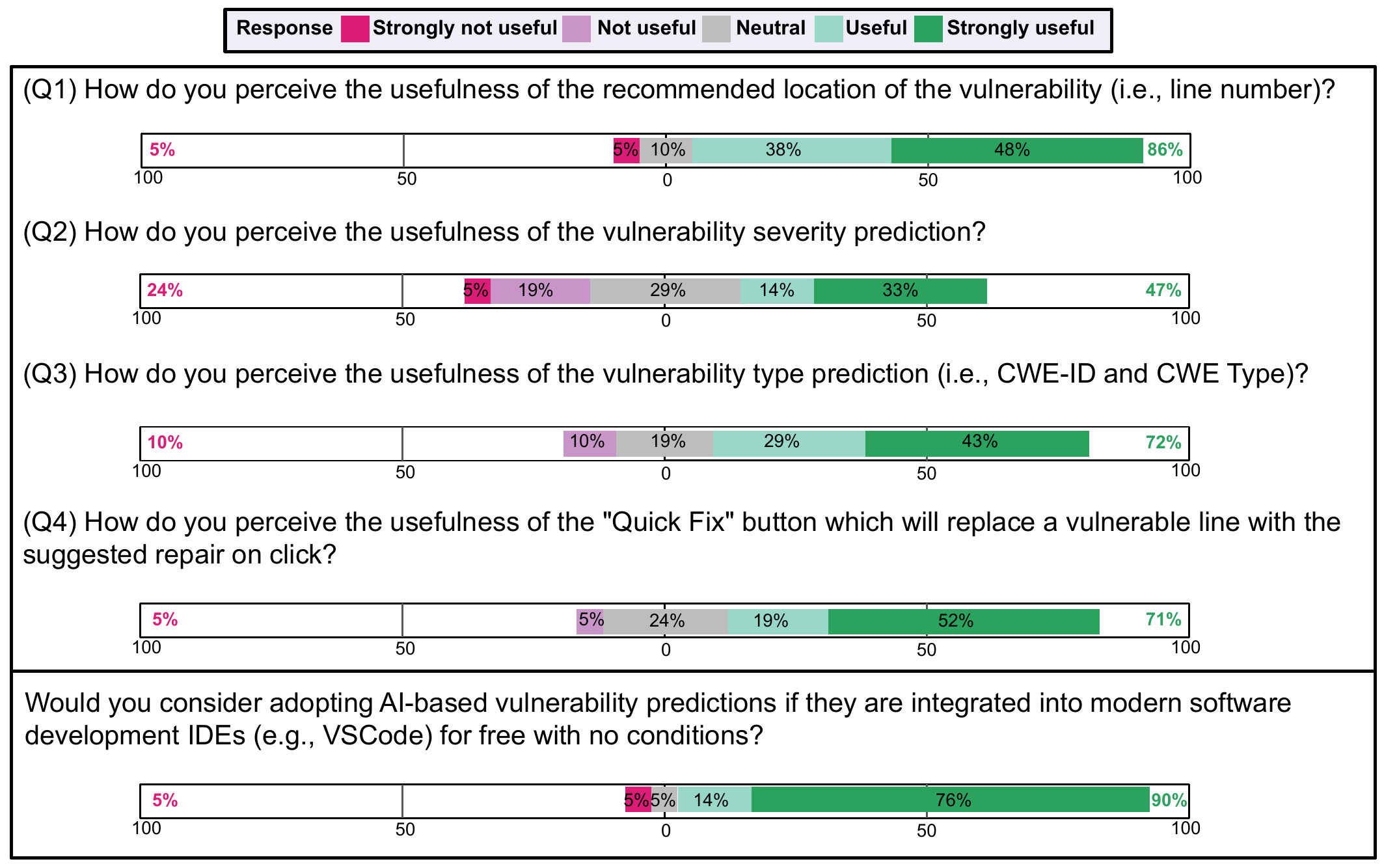}
\caption{(Q1-Q4) A summary of the survey questions and the results obtained from 21 participants.}
\label{fig:survey}
\end{figure*}

\subsubsection{Survey Results}
Fig~\ref{fig:survey} summarizes the survey results, we describe each question in detail in the following.

\begin{figure*}[t]
\includegraphics[width=\linewidth]{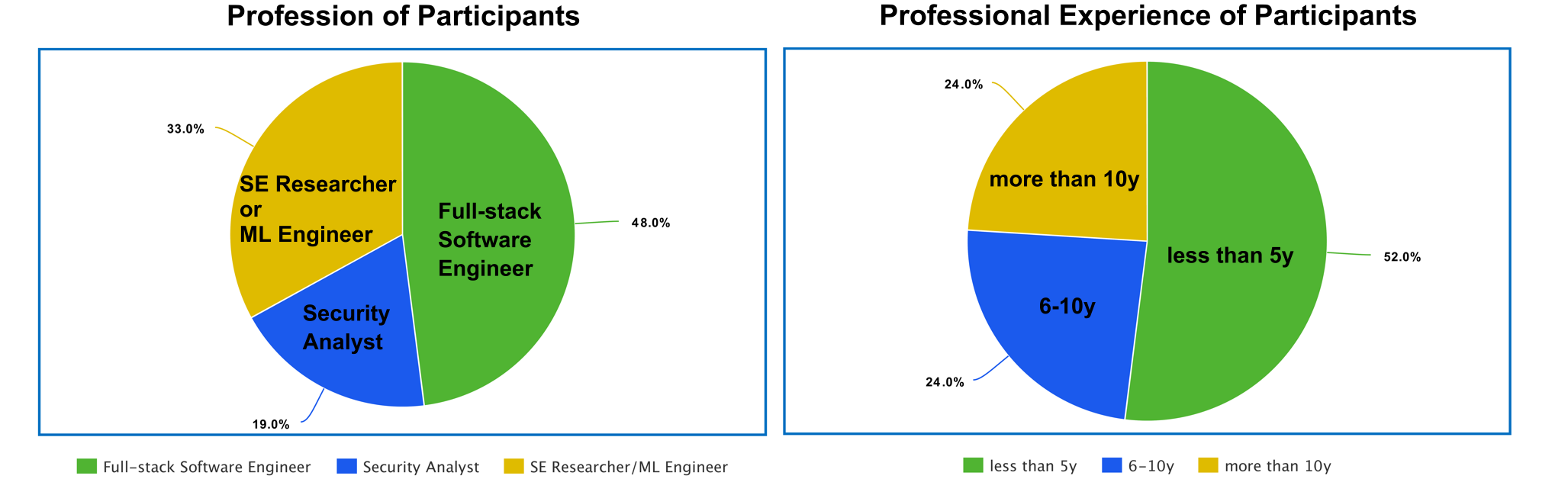}
\caption{The demographics of our survey participants in terms of their profession and professional experience.}
\label{fig:demographic}
\end{figure*}

\smallsection{Respondent demographics}
Fig~\ref{fig:demographic} presents the overall respondent demographic.
In terms of the profession of the participants, 48\% ($\frac{10}{21}$) of them are full-stack software engineers, 19\% ($\frac{4}{21}$) of them are security analysts, while the other 33\% are software engineering researchers, machine learning engineers, etc.
In terms of the level of their professional experience, 
52\% ($\frac{11}{21}$) of them have less than 5 years of experience, 24\% ($\frac{5}{21}$) have 6-10 years of experience, while the other 24\% have more than 10 years of experience.

\noindent\textbf{(Q1) How do you perceive the usefulness of the recommended location of the vulnerability (i.e., line number)?} 

\noindent \smallsection{Findings} \textbf{86\% of the respondents perceived that the prediction of the vulnerability location is useful} due to various reasons:

\begin{itemize}
    \item Explicitly localize the vulnerability (R1: \textit{I think it is useful to know which line does the vulnerability locate.}, R7: \textit{It can help you quickly identify where the vulnerability is.}, R15: \textit{It is helpful to know the reason for vulnerability and line number.})
    \item Reduce time spent on code review (R4: \textit{Speed to fix for developers.}, R11: \textit{This would decrease code review time when I want to check for security breaches.}, R12: \textit{This is something it would take me a lot of time debugging to figure out.})
    \item Support debugging process (R6: \textit{Been useful in debugging code}, R9: \textit{Helps with quick resolution of bugs/vulnerabilities.})
\end{itemize}

\noindent\textbf{(Q2) How do you perceive the usefulness of the vulnerability severity prediction?}

\noindent \smallsection{Findings} \textbf{47\% of the respondents perceived that the prediction of the vulnerability severity score is useful} due to various reasons:

\begin{itemize}
    \item Prioritization of vulnerability repairs (R3: \textit{Having it will allow me to prioritize fixing high-impact vulnerabilities before looking at things that don't matter as much.}, R8: \textit{Vulnerability score will help me prioritize the fix.}, R11: \textit{This would help me prioritize which part of the code I should fix first.})
    \item Risk management (R4: \textit{Modeling business risk is very useful for overall software development planning.}, R9: \textit{Determines the magnitude of the vulnerability against the risks involved.}, R13: \textit{Just like vulnerability scanning tools, it is helpful to know how bad it is and decide the further steps, so yes, it is useful.})
\end{itemize}

\noindent\textbf{(Q3) How do you perceive the usefulness of the vulnerability type prediction (i.e., CWE-ID and CWE-Type)?} 

\noindent \smallsection{Findings} \textbf{72\% of the respondents perceived that the predictions of CWE-ID and CWE-Type are useful} due to various reasons:

\begin{itemize}
    \item Help understands the vulnerability (R3: \textit{It is important to understand what the vulnerability is before you fix it.}, R9: \textit{Helps to identify common weaknesses and resolve them easily.}, R11: \textit{This would help me understand which kind of security breach I might be facing.}, R15: \textit{It helps understand the problem.}, R19: \textit{Easier to know what the problem is.})
    \item Align with security practices (R4: \textit{Aligns well with security practices.}, R13: \textit{A quick classification would help solve the problem more efficiently.})
\end{itemize}

\noindent\textbf{(Q4) How do you perceive the usefulness of the ``Quick Fix'' button which will replace a vulnerable line with the suggested repair on click?} 

\noindent \smallsection{Findings} \textbf{71\% of the respondents perceived that the ``Quick Fix'' button that suggests the vulnerability repair is useful} due to various reasons:

\begin{itemize}
    \item Reduce time spent on vulnerability repairs (R6, R15: \textit{It saves time.}, R9: \textit{Save hours to resolve an issue.}, R11: \textit{This would save a lot of time. I can also modify the suggested codes if I want to.})
    \item Help with vulnerability repair implementation (R12: \textit{Having a potential fix helps me think through the fix I would like to implement even if I do not use the suggested fix.})
\end{itemize}

\smallsection{Summary} Our survey study with 21 software practitioners found that all kinds of vulnerability predictions provided by our \ourtool~are perceived as useful. Specifically, the vulnerable line prediction reduces the time required to locate vulnerability while severity score prediction helps developers prioritize their workloads. Moreover, the prediction of the vulnerability type helps developers understand the vulnerability and the repair recommendation suggested by the ``Quick Fix'' button helps developers come up with repair implementation. Finally, we found that \textbf{90\% of the respondents consider adopting an AI-based vulnerability prediction approach such as \ourtool}~if it is publicly available for free in a modern IDE (e.g., Visual Studio Code), highlighting the practical need for our AI-based vulnerability prediction approach.

\subsection{A Preliminary User Study}
\begin{figure*}[t]
\includegraphics[width=\linewidth]{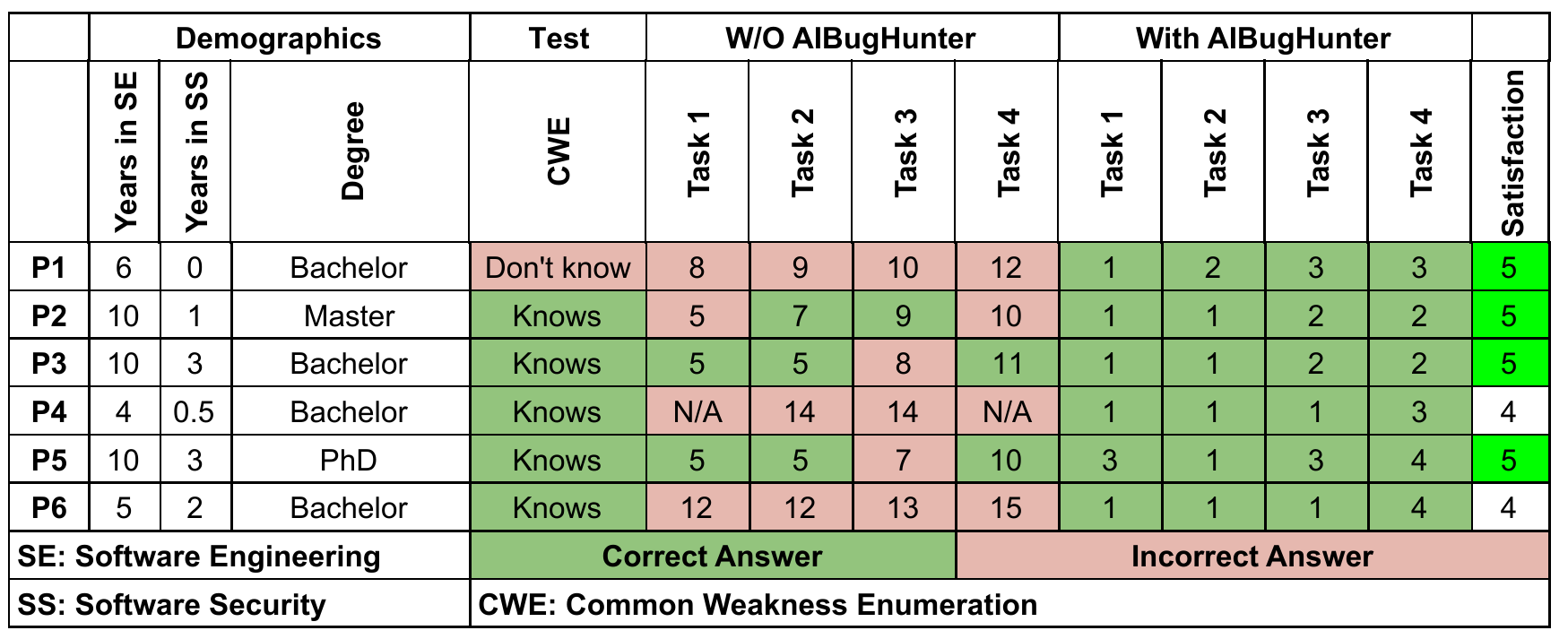}
\caption{The experimental results of our user study with six participants. Wherein the first task was to locate the vulnerability, the second task was to explain the vulnerability type, the third task was to estimate the vulnerability severity, and the fourth task was to suggest repairs. The time was measured in minutes and the satisfaction ranges from 1 (highly dissatisfied) to 5 (highly satisfied).}
\label{fig:exp_results}
\end{figure*}

We conducted a preliminary user study to assess the impact that \ourtool~may have on developers' productivity in security analysis.
To do so, we choose \emph{single-subject experimental designs} as our research methodology---a type of research methodology characterized by repeated assessment of a particular phenomenon (often a behavior) over time.
The single-subject experiment is useful when researchers are attempting to observe the behavior of an individual or a small group of individuals and wishes to document that observation.
In particular, we run the user study with two groups, i.e., a control group (i.e., the group of participants that do not have access to \ourtool) and a treatment group (i.e., the group of participants that have access to \ourtool).
First, we assign a vulnerable C/C++ function to a participant to perform given tasks.
The tasks are to locate, estimate severity, explain its type, and suggest repairs.
In a well-designed experiment, all variables apart from the treatment should be kept constant between the two groups, allowing us to correctly measure the entire effect of the treatment without interference from confounding variables.
With this methodology, our results will not be affected by different participants' expertise and task difficulty (which is commonly affected by randomized control trials).
In what follows, we illustrate our user study design followed by the results.

\subsubsection{User Study Design}
\textbf{(Step 1) Design and develop a user study.}
Our user study is face-to-face where each participant participated individually.
Our user study consists of three parts, (1) demographic questions, (2) user study, and (3) survey questions to seek feedback after using \ourtool.

In the demographic questions, we asked about the participants' education and experience in software engineering and software security to ensure that we approach the right target group of participants.

In the user study, we used a real-world vulnerable C function in our experiment~\citep{experiment}. The \textit{jpeg\_size} function from a PDF generation library caused a buffer overread vulnerability (i.e., CWE-125) due to an inappropriate data bounding check.
In particular, we designed our main experiment into two parts.
The participants were asked to diagnose the vulnerable C function without using our \ourtool~tool in the first part while they were asked to diagnose the vulnerable function with the help of our \ourtool~tool in the second part.
In each part, the participants were required to complete four tasks within 15 minutes, i.e., (1) locate vulnerability, (2) explain the vulnerability type, (3) estimate vulnerability severity, and (4) suggest repairs.

In the survey questions, we asked about the participants' satisfaction with our \ourtool~using a Likert scale ranging from 1 to 5 followed by an open-ended question for justification.
Last but not least, our experiment has been rigorously reviewed and approved by the Monash University Human Research Ethics Committee (MUHREC ID: 36037).

\textbf{(Step 2) Recruit and select participants.}
We recruited software developers and researchers that have software engineering and/or software security expertise.
To ensure the diversity of our participants, we select participants from a diverse set of  professional experiences and occupations.
Finally, we recruited a total of 6 participants to participate in our user study.
Each participant will receive a gift card of \$20 as a token of appreciation.

\textbf{(Step 3) Conduct the user study.}
We conducted the user study as mentioned in Step 1. 
We also video-recorded during the user study with permission from the participants.
Finally, for each participant, we analyzed the time spent on each task between the two groups (i.e., control vs treatment).  
Then, we manually analyzed the responses to the open-ended questions to better understand the in-depth insights from the participants.

\subsubsection{User Study Results}
\noindent{\textbf{\underline{Participant Demographics.}}}
The education level of our participants varies from bachelor, master, to Ph.D. degrees, while the professional experience in software engineering and software security varies from a few months to 10 years, ensuring that the results are not bounded to specific groups of participants.

\noindent{\textbf{\underline{Main Findings.}}}
\textbf{Our \ourtool~can reduce the time spent on detecting, locating, estimating, explaining, and repairing vulnerabilities from 10-15 minutes to 3-4 minutes (see Figure~\ref{fig:exp_results})}.
Without using \ourtool, the results show that the majority of the participants cannot provide accurate answers to the given tasks, which indicates that the vulnerability analysis task is challenging and time-consuming.
With the use of \ourtool, the results show that all of the participants were able to provide accurate answers to the given tasks within 4 minutes.
This finding implies that \ourtool~could possibly enhance developers' productivity in combating cybersecurity issues during the software development lifecycle. Last but not least, all of the participants rated our \ourtool~as satisfied or highly satisfied due to reasons as follows:
\begin{itemize}
    \item P1: \emph{It is seamlessly integrated into my development environment.}
    \item P3: \emph{It exceeds my expectations for automated tools.}
    \item P4: \emph{Detect the vulnerability down to line-level and provide CWE information.}
    \item P5: \emph{Identify the vulnerability fast.}
\end{itemize}

\subsection{The implications of \ourtool~to researchers and practitioners}
In this section, we discuss the broader implications of our \ourtool~tool for researchers and practitioners. For practitioners, our \ourtool~tool can help security practitioners locate vulnerabilities, identify vulnerability types, estimate vulnerability severity, and suggest vulnerability repairs. These AI-powered security intelligence features can produce significant benefits to practitioners. This includes potentially increasing developers’ productivity, increasing the security of their software systems, and reducing overall software development costs. For researchers, our \ourtool~tool is among the first proof-of-concept AI-powered security intelligence tool with numerous features combined into one tool. Many static analysis tools can only perform vulnerability detection, not repairs. Instead, we present how such important features could be integrated into a VS Code Extension. The results of our user study also highlight the usability of our tool and its substantial potential benefits for the software engineering community.
\section{Threats to Validity}
\label{sec:threats}

\subsection{Construct Validity}
\textbf{Threats to the construct validity} relate to the potential bias of our survey study and user study. In our survey study and user study, we recruited 21 and 6 participants respectively from different professions such as software engineers and security analysts. However, the results of our two studies could still be biased towards our participants and the results do not necessarily generalize to other audiences. To mitigate this threat for our survey study, we spread our survey on social platforms such as Facebook and LinkedIn to ensure diverse participant demographics. To mitigate this threat for our user study, we recruited software practitioners with different backgrounds and professional experiences for our user study.

The goal of our survey study and user study is to investigate the usefulness of the tool. Thus, we only focus on correct predictions when designing our survey study and user study. However, our \ourtool~could also return incorrect predictions. Thus, an extended user study is also required to fully evaluate the impact of our \ourtool~by including both correct and incorrect predictions. Since this research question requires a rigorous user study and a different methodology than we use in this article, we plan to investigate this in future work.

Furthermore, the maturity of \ourtool~is still at the early stage of development and is not yet ready for commercialization.
Our user study experiment was conducted as a preliminary analysis.
Thus, the findings are only limited to our studied group, and may not be generalized to other participants, users, software systems, and organizations.
Therefore, an extensive evaluation of \ourtool~is still needed.

\subsection{Internal Validity}
\textbf{Threats to the internal validity} relate to our choice of hyperparameter settings (i.e., optimizer, scheduler, learning rate, etc.) of our models to classify vulnerability types and estimate vulnerability severity.
Finding a set of optimal hyperparameter settings of the CodeBERT model is extremely expensive due to a large number of trainable parameters in CodeBERT and the large search space of the Transformer architecture.
Thus, we leverage the default setting of CodeBERT as reported by Feng~\ea~\cite{feng2020codebert}.
Hence, our results serve as a lower bound for our approach, which can be further improved through hyperparameter optimization~\cite{tantithamthavorn2016automated,tantithamthavorn2018optimization}. To mitigate this threat, we report the hyperparameter settings in the replication package to support future replication studies.

\subsection{External Validity}
\textbf{Threats to the external validity} relate to the generalizability and applicability of our \ourtool. The models used in \ourtool~were trained using Big-Vul~\cite{fan2020ac} and CVEFixes~\cite{bhandari2021cvefixes} datasets consisting of C/C++ source code. Thus, our models do not necessarily generalize to other data and programming languages. However, the \ourtool~tool could be used with other programming languages as it is designed to adopt any deep learning models. Nevertheless, future work could explore the effectiveness of the \ourtool~tool in other programming languages when other models are used.

\section{Related Work}
\label{sec:related}
We discuss key previous studies of ML-based vulnerability prediction and multi-task learning for software vulnerability prediction. We compare our approach with previous methods and illustrate the difference.

\subsection{ML-Based Vulnerability Type Classification}
Multiple ML-based approaches have been proposed to automate the CWE-ID classification task~\citep{na2016study, aota2020automation, shuai2013automatic}.
\cite{shuai2013automatic} constructed a Huffman Tree SVM
, \cite{na2016study} used a Naive Bayes model
, and~\cite{aota2020automation} leveraged a Random Forest model to automate the CWE-ID classification task.
All of these approaches rely on the Bag of Words technique, 
while such a method can embed textual input features into numeric vector space,
such embedding based on word counting can not capture enough semantic information of input.

Instead of using CVE entries as input,
\cite{wang2020machine} leveraged ML-based models to classify CWE-IDs for vulnerability security patches based on the features extracted from security patches. 
However, defining such hand-crafted features is time-consuming and may require much effort.

Recently, researchers have proposed DL-based models that learn the input representation through neural networks to better capture the semantic features of the input.
\cite{aghaei2020threatzoom} proposed ThreatZoom, a Hierarchical Neural Network that considers the hierarchical nature of CWE-ID.
\cite{wang2021towards} leveraged the BERT architecture to learn textual features through the self-attention mechanism.

Previous studies focus on mapping either CVE entries (i.e., vulnerability description) or security patches into CWE-ID, however, such input features are not available during the software development stage, thus they are not compatible with our \ourtool, where it requires an ML model to predict based on the source code written by developers.
In contrast, our approach only takes vulnerable source code without any description as input and predicts the corresponding CWE-ID.
Therefore, it can support our \ourtool~to generate vulnerability predictions based on the code written by developers.

\subsection{Multi-Task Learning for Software Vulnerability Prediction}
\cite{spanos2018multi} used three ML ensemble classifiers to predict CVSS characteristics based on vulnerability description.
\cite{le2021deepcva} proposed DeepCVA which uses multiple GRUs and a shared embedding layer as a multi-task learning framework for commit-level vulnerability assessment.
\cite{gong2019joint} leveraged a Bi-LSTM as a shared feature extractor with multiple classifiers to predict different Common Vulnerability Scoring System (CVSS) characteristics based on vulnerability description.
\cite{babalau2021severity} used a shared BERT architecture with two prediction heads to learn a multi-task model which supports CVSS severity score classification and regression.

Some of these studies leveraged a shared architecture~\citep{gong2019joint, babalau2021severity,takerngsaksiri2022syntax} that can learn from labels of different tasks that are correlated, hence may help improve the model performance.
Nevertheless, all of these studies relied on the weighted summation of loss functions during gradient descent, i.e.,
(1) averaging the loss of each task~\citep{le2021deepcva}, 
(2) tuning loss weights of each task~\citep{babalau2021severity},
(3) summarizing loss of each task~\citep{gong2019joint}.
Such a weighted summation approach may not find the optimal solution when updating the shared model, for instance, the updated parameters are better for one task but not the other as discussed in Section~\ref{sec:movul_moo}.
In contrast, our approach finds an optimal collection of parameters that benefits all tasks simultaneously during gradient descent that can optimize a collection of possibly conflicting objectives.
To the best of our knowledge, 
this paper is among the first to leverage multi-objective optimization to learn a DL model for the software vulnerability classification task.

\subsection{Explainable AI for Cybersecurity}

Explainability is now becoming a critical concern in software engineering.
Many researchers often employed AI/ML techniques for defect prediction~\citep{pornprasit2021jitline,pornprasitpyexplainer,pornprasit2022deeplinedp,wattanakriengkrai2020predicting,rajapaksha2021sqaplanner}, malware detection~\citep{DBLP:conf/issre/LiuT0022,DBLP:journals/csur/LiuTLL23}, and effort estimation~\citep{fu2022gpt2sp}.
Yet, little is focused on explaining the vulnerability predictions, which is the focus of this paper.
While these AI/ML techniques can greatly improve developers' productivity, software quality, and end-user experience, practitioners still do not understand why such AI/ML models made those predictions \citep{tantithamthavorn2021actionable,tantithamthavorn2020explainable,jiarpakdee2021perception,jiarpakdee2020modelagnostic,rajapaksha2021sqaplanner,pornprasitpyexplainer}.
In particular, the survey study by~\cite{jiarpakdee2021perception} found that explaining the predictions is as equally important and useful as improving the accuracy of defect prediction.
However, their literature review found that 91\% (81/96) of the defect prediction studies only focus on improving the predictive accuracy, without considering explaining the predictions, while only 4\% of these 96 studies focus on explaining the predictions.

Although Explainable AI is still a very under-researched topic within the software engineering community~\citep{tantithamthavorn2021explainable,tantithamthavorn2021actionable,10109341,10109328}, very few existing XAI studies have shown some successful usages e.g., in defect prediction. In one example, \cite{wattanakriengkrai2020predicting} and \cite{pornprasit2021jitline} employed model-agnostic techniques (e.g., LIME) for line-level defect prediction (e.g., predicting which lines will be defective in the future), helping developers to localize defective lines in a cost-effective manner.
For example, \cite{jiarpakdee2020modelagnostic} and \cite{khanan2020jitbot} employed model-agnostic techniques (e.g., LIME) for explaining defect prediction models, helping  developers better understand why a file is predicted as defective.
\cite{rajapaksha2021sqaplanner} and \cite{pornprasitpyexplainer} proposed local rule-based model-agnostic techniques to generate actionable guidance to help managers chart the most effective quality improvement plans.

In contrast to the prior studies, but in the same vein of research in explainable AI for software engineering, this paper aims to go beyond vulnerability prediction but provides explanations on the types, the severity, and the suggested repairs. 

\section{Conclusions}
\label{sec:conclusion}
\revisedinline{In this article, we propose \ourtool, an integration of our proposed software vulnerability classification (multi-objective optimization approach) and estimation (a transformer-based approach) approaches and our previous works. Our \ourtool~is an ML-based vulnerability prediction tool to (1) localize vulnerabilities, (2) classify vulnerability types, (3) estimate vulnerability severity, and (4) suggest repairs.}
To the best of our knowledge, this article is among the first to deploy an ML-based vulnerability prediction tool for C/C++ to the VS Code IDE.
Our \ourtool~realizes real-time vulnerability prediction during software development, which helps integrate security approaches into the software development life cycle.
\revisedinline{Our empirical survey study with 21 software practitioners confirms that our \ourtool~is perceived as useful;
and our user study indicates that our \ourtool~could help reduce developers' time spent on security analysis, which could enhance developers’ productivity in combating security issues during software development.}

\textbf{Data Availability Statement}
The data, model training and evaluation scripts that support the findings of this study are available at: (\url{https://github.com/awsm-research/AIBugHunter}).
Our proposed \ourtool~is available at: (\url{https://marketplace.visualstudio.com/items?itemName=AIBugHunter.aibughunter}).

\textbf{Conflict of Interest Statement}
The authors of this article declared that they have no conflict of interest.

\begin{acknowledgements}
Chakkrit Tantithamthavorn was partly supported by the Australian Research Council's Discovery Early Career Researcher Award (DECRA) (DE200100941). 
John Grundy is supported by ARC Laureate Fellowship (FL190100035).
\end{acknowledgements}

\bibliographystyle{spbasic}
\bibliography{myref,additionalref}

\end{document}